\documentclass{article}

\usepackage{arxiv}

\usepackage[utf8]{inputenc} 
\usepackage[T1]{fontenc}    
\usepackage{hyperref}       
\usepackage{url}            
\usepackage{booktabs}       
\usepackage{amsfonts}       
\usepackage{nicefrac}       
\usepackage{microtype}      
\usepackage{graphicx}
\usepackage{natbib}
\usepackage{doi}
\usepackage{amsmath}
\usepackage{amssymb}
\usepackage{multirow}

\title{From Sea Surface Height to Velocity - A machine learning approach}

\author{
  Yaswanth Reddy Yaradoddi \\
  Department of Computer Science\\
  Texas A\&M University \\
  yaswanthreddy@tamu.edu
  \And
  Jinbo Wang \\
  Department of Oceanography\\
  Texas A\&M University \\
  jinbo.wang@tamu.edu\\
  \And
  Elyes Serghine \\
  Computer Science\\
  California Institute of Technology \\
  eserghin@alumni.caltech.edu
}

\renewcommand{\shorttitle}{SSH to Velocity via ML}

\hypersetup{
pdftitle={From Sea Surface Height to Velocity: A Machine Learning Approach},
pdfsubject={Physical Oceanography, Machine Learning},
pdfauthor={Yaswanth Reddy Yaradoddi, Jinbo Wang, Elyes Serghine},
pdfkeywords={sea surface height, ocean velocity, U-Net, LLC4320, geostrophic balance, machine learning},
}

\begin{document}
\maketitle

\begin{abstract}
Inferring upper-ocean velocity from altimetric sea surface height (SSH) is a fundamental problem in satellite oceanography, but an unsolved problem for the scales below approximately 50 km that are now observable because of the Surface Water and Ocean Topography (SWOT) mission but not before. The conventional methods are fundamentally limited by assumed balance dynamics either from geostrophy or cyclogeostrophy. Here, we use a deep-learning model trained on a high-resolution ocean simulation at 1.5 km grid space to demonstrate that SSH, with supporting environment data, can predict the full velocity vector and can even recover diagnostics that require spatial derivatives such as spectral energy flux.  A simple U-Net encoder--decoder was trained on hourly fields from the pre-SWOT ``Cape Basin'' extraction of the LLC4320 ocean simulation (11.8--15.8$^\circ$E, 41--45$^\circ$S), with the horizontal velocity as the target and SSH, sea surface temperature (SST), and wind stress as inputs. We tested the prediction of velocity at surface and 50 meters where winds have less influence.  For the 50-m velocity prediction, the best model has a RMSE of speed 0.049\,m/s, 29\,\% below geostrophic approximation, and reproduces relative vorticity with correlation of 0.93, compared with 0.75 for geostrophic approximation. Importantly, our model can produce divergence with correlation at 0.66, while geostrophic approximation has zero divergence by definition. We have tested many different strategies in training the model as summarized in this paper. The four most useful observations are 1. SSH gradients are better inputs than raw SSH, 2. wind stress is the most useful covariate for total velocity prediction, 3. the benefit of including SST in the input varies seasonally, 4. including vorticity and divergence in the loss improves the prediction accuracy. We use this study to show a path to transfer high-fidelity physics in simulations towards applications on SWOT  that produces snapshots of SSH but with km-scale high spatial resolution, where we have observed superposition of balanced and unbalanced SSH.
\end{abstract}


\section{Introduction}

Satellite altimetry has transformed physical oceanography by providing global observations of sea surface height (SSH), from which the large-scale and mesoscale surface geostrophic circulation can be estimated. Under hydrostatic and geostrophic balance, the horizontal pressure-gradient force associated with sea-surface dynamic topography $\eta$ is balanced by the Coriolis acceleration, yielding the familiar diagnostic relation
\begin{equation}
    \mathbf{u}_g
    =
    \frac{g}{f}\,\hat{\mathbf{z}}\times\nabla \eta
    =
    \left(
    -\frac{g}{f}\frac{\partial \eta}{\partial y},
     \frac{g}{f}\frac{\partial \eta}{\partial x}
    \right),
    \label{eq:geostrophic_velocity}
\end{equation}
where $\mathbf{u}_g=(u_g,v_g)$ is the geostrophic velocity, $g$ is gravitational acceleration, $f$ is the Coriolis parameter, and $\hat{\mathbf{z}}$ is the local vertical unit vector. This linear relationship has been the basis of many altimetry-derived surface-current products and remains highly effective when the Rossby number $Ro=U/(fL)$ is small and the flow evolves slowly compared with the local inertial period.

The assumptions behind Eq.~\eqref{eq:geostrophic_velocity}, however, become increasingly restrictive at the spatial scales now accessible to wide-swath altimetry. The Surface Water and Ocean Topography (SWOT) mission provides two-dimensional SSH observations at unprecedented resolution, exposing sea-level variability at scales where ageostrophic motions, internal waves, and measurement noise are no longer secondary \citep{Archer2025}. As the horizontal scale decreases toward $\mathcal{O}(10$--$100)$ km, $Ro$ can approach order one in energetic fronts, eddies, western boundary currents, and equatorial regions. In this regime, nonlinear advection contributes at leading order to the momentum balance, and the mapping from SSH to velocity is no longer a simple local rotation and rescaling of the SSH gradient. The difficulty is particularly acute near the equator, where $f\rightarrow 0$ makes Eq.~\eqref{eq:geostrophic_velocity} singular and amplifies SSH-gradient errors; consequently, altimetric geostrophic velocities are often masked or treated with special equatorial dynamics, while independent constraints from drifters or other observing systems become essential.

A useful way to see the limitation of geostrophy is to begin with the horizontal momentum equation,
\begin{equation}
    \frac{\partial \mathbf{u}}{\partial t}
    + \mathbf{u}\cdot\nabla\mathbf{u}
    + f\hat{\mathbf{z}}\times\mathbf{u}
    =
    -g\nabla \eta
    + \mathbf{r},
    \label{eq:momentum_full}
\end{equation}
where $\mathbf{r}$ denotes residual effects such as friction, unresolved vertical structure, and other non-conservative forces. If the tendency and residual terms are neglected but nonlinear advection is retained, Eq.~\eqref{eq:momentum_full} becomes the steady nonlinear balance
\begin{equation}
    \mathbf{u}\cdot\nabla\mathbf{u}
    + f\hat{\mathbf{z}}\times\mathbf{u}
    =
    -g\nabla \eta .
    \label{eq:steady_nonlinear_balance}
\end{equation}
Writing the velocity in natural coordinates as $\mathbf{u}=V\hat{\mathbf{s}}$, with $\hat{\mathbf{s}}$ tangent to the streamline and $\hat{\mathbf{n}}$ the signed left-normal direction, gives
\begin{align}
    V\frac{\partial V}{\partial s}
    &= -g\frac{\partial \eta}{\partial s},
    \label{eq:natural_tangential}\\
    \kappa V^2 + fV
    &= -g\frac{\partial \eta}{\partial n},
    \label{eq:natural_normal}
\end{align}
where $V$ is speed and $\kappa$ is the signed streamline curvature. Eq.~\eqref{eq:natural_normal} is the gradient-wind (cyclogeostrophic) balance among the centripetal acceleration, Coriolis acceleration, and pressure-gradient force; if the flow is locally axisymmetric or slowly varying along streamlines, $\partial V/\partial s\approx 0$ and, by Eq.~\eqref{eq:natural_tangential}, the flow follows SSH contours. The geostrophic relation is recovered when the curvature term $\kappa V^2$ is small. In contrast, when $Ro$ is finite, the curvature term can be comparable to the Coriolis term; when $f$ is small, the pressure gradient may instead balance the centripetal acceleration in a cyclostrophic limit. Thus, even within balanced dynamics, SSH gradients alone do not uniquely determine the velocity without additional information about streamline geometry, flow direction, and the appropriate dynamical regime.

The problem becomes more underdetermined when unbalanced motions are present. Linear and finite-amplitude internal gravity waves, internal tides, near-inertial motions, and surface-wave-related signals can all project onto SSH at wavelengths that overlap with balanced mesoscale and submesoscale motions \citep{Archer2025,kachelein2025,zhang.callies2025}. For these motions, the tendency term in Eq.~\eqref{eq:momentum_full} is not a small residual; it is part of the leading-order balance. A single satellite snapshot provides $\eta(x,y,t_0)$ but not the temporal tendency, wave phase propagation, vertical-mode structure, or the partition between balanced and unbalanced components. Consequently, two physically different velocity fields can be associated with very similar SSH snapshots. This non-uniqueness cannot be eliminated by a more flexible regression model alone; it can only be regularized through physical assumptions, statistical priors, training data, or additional observations.

Machine learning offers a promising framework for such regularization. Recent studies have used neural networks and related methods to infer surface currents or surface kinematic quantities from satellite-observable variables, reconstruct vorticity, strain, and divergence from SSH snapshots, separate balanced and unbalanced SSH components, or infer SWOT sea-surface velocity from collocated drifter observations \citep{Sinha2021,Xiao2023,Gao2024,Zhou2025,martin2026generative}. These studies demonstrate that data-driven models can exploit nonlocal spatial structure, learn approximate dynamical closures, and improve upon purely geostrophic estimates in some regimes. However, the success of machine learning does not by itself establish that velocity is identifiable from SSH alone. Model skill may depend on the training distribution, the amplitude of internal waves, latitude, Rossby number, SSH noise, the target definition of velocity, and whether the validation data share the same balance assumptions as the training data.

The present study addresses these questions with a U-Net trained on the LLC4320 global ocean simulation \citep{Rocha2016,Gallmeier2023}. Our main target is the horizontal velocity at model level~16 (cell center at 52.7\,m depth), hereafter $(u,v)$; we also train the same configuration for the surface velocity. This paper makes the following contributions:
\begin{enumerate}
    \item Training on one period and evaluating on a later, held-out period, with three random seeds per configuration, we show that a standard U-Net is no better than geostrophy, and we identify the training choices that make it generalize: rotation and crop augmentation, predicting the departure from geostrophy, vorticity and divergence penalties, and moderate capacity.
    \item We revisit the questions of input representation (raw SSH versus SSH gradients), auxiliary covariates (SST and wind stress), capacity, and training objective for the final configuration.
    \item We evaluate the physical consistency of the predictions through relative vorticity, horizontal divergence, and the joint distribution of vorticity and strain, and compare them with $f$-plane geostrophy.
    \item We compare the spectral kinetic-energy flux of the true, geostrophic, and predicted velocities at 53\,m and at the surface, and show that geostrophy misrepresents the cascade of energy across scales whereas the U-Net reproduces it.
\end{enumerate}

The remainder of the paper is organized as follows. Section~\ref{sec:previous_studies} reviews previous machine-learning approaches. Section~\ref{sec:architecture} describes the data, the training and evaluation periods, the model, and the training setup. Section~\ref{sec:training_results} builds a model that generalizes and presents the ablations. Section~\ref{sec:validation_results} examines the joint statistics of vorticity and strain, Section~\ref{sec:flux} the spectral energy flux, Section~\ref{sec:discussion} discusses the results, and Section~\ref{sec:conclusion} concludes.

\section{Previous studies}
\label{sec:previous_studies}

Estimating ocean velocity from sea surface height is a long-standing diagnostic problem, but the meaning of the target velocity differs substantially across studies. Classical altimetric products typically estimate the geostrophic component of the surface circulation from gradients of dynamic topography. This approach is physically transparent and robust at large scales, but it implicitly assumes small Rossby number, weak time dependence, and a leading-order geostrophic momentum balance. These assumptions become less reliable at smaller spatial scales, in energetic fronts and eddies, in low-latitude regions where $f$ is small, and in the presence of internal waves, tides, and other unbalanced motions. Recent machine-learning studies have therefore explored whether statistical models can use the spatial structure of SSH, together with other observable variables or learned physical priors, to infer velocity or related dynamical quantities beyond the purely geostrophic approximation.

A useful first step was provided by \citet{Sinha2021}, who framed surface-current estimation as a supervised learning problem using satellite-observable variables. Their models were trained on daily surface fields of a global $\approx0.1^\circ$ primitive-equation ocean simulation (CESM-POP), with inputs including SSH, sea-surface temperature and salinity, wind stress, and transformed geographical coordinates, and with the model surface velocity as the target. They showed that a neural network can predict surface currents over most of the global ocean with lower error than a geostrophy-plus-Ekman estimate, particularly when given a local stencil of neighboring SSH values from which it can learn spatial gradients. This result is important because it demonstrates that data-driven models can learn more than a pointwise balance approximation: they can exploit spatial gradients, latitude-dependent dynamics, and empirical relationships between wind forcing, temperature structure, and surface currents. At the same time, the study also highlights a key limitation for the present problem. The successful models were not constrained to use SSH alone; they used additional physical predictors and were trained and evaluated in a fully sampled, noise-free model environment. Their errors were largest where the local Rossby number is high, suggesting that the mapping from surface height to total velocity remains difficult precisely in the regimes most relevant to SWOT-scale variability.

A related line of work has focused not on the velocity vector itself, but on surface kinematic quantities that are dynamically informative at the submesoscale. \citet{Xiao2023} used convolutional neural networks to reconstruct surface vorticity, strain, and divergence from snapshots of SSH in submesoscale-permitting simulations, including an idealized channel model and regions of LLC4320. This problem is closely connected to velocity inference because vorticity and strain involve spatial derivatives of the velocity field, and are therefore even more sensitive to small-scale SSH structure and noise than velocity itself. Their results showed that neural networks can recover vorticity and strain with useful skill when balanced dynamics dominate and inertial-gravity-wave amplitudes are relatively weak. They also found that divergence was a surprisingly reliable field to reconstruct, because the network naturally filters the wave-driven divergence, whereas strong internal-wave signals degrade the reconstruction of vorticity and strain. Thus, this work supports the idea that SSH snapshots contain nonlocal information about surface dynamics, but it also reinforces the point that the learned mapping depends on the relative amplitudes of balanced and unbalanced motions in the training distribution. Our work is different from \citet{Xiao2023} by directly predicting velocity vector instead of predicting vorticity/strain/divergence, which are derived quantities of velocity vector. 

At the global scale, \citet{Martin2024} showed that a deep-learning SSH mapping system (NeurOST), trained only on real along-track altimetry with SST as an additional input, resolves finer eddy structure than the standard DUACS product. In an observing-system simulation experiment (OSSE) it recovers 50--80\% of the surface vorticity variance in the subtropics and western boundary currents, and applied to real observations it reveals a strongly seasonal, spring-peaking kinetic-energy cascade whose seasonality is largely absent in DUACS. Their evaluation of vorticity maps against a conventional product and of vorticity skill in an OSSE motivates the kinematic evaluation we adopt in Section~\ref{sec:kinematic_metrics}. More recently, \citet{Lyu2024} used a U-Net to decompose instantaneous LLC4320 SSH snapshots from the Agulhas region into balanced and unbalanced components, and showed that input whitening and data augmentation improve the decomposition across the wavenumber spectrum and in SSH gradients, relative to Gaussian filtering; \citet{Wang2026} extended this approach to a probabilistic decomposition that also estimates the uncertainty of the balanced and unbalanced components.

Another closely related problem is the separation of balanced and unbalanced SSH signals. \citet{Gao2024} developed a deep-learning approach to extract balanced motions from a single SSH snapshot, motivated by the fact that wide-swath altimetry provides high spatial resolution but relatively sparse temporal sampling. Their method uses realistic simulation data to learn a decomposition between balanced geostrophic motions and unbalanced wave motions, and was shown to retain gradient and Laplacian information more effectively than simple spatial filtering. This is particularly relevant for SWOT because balanced and unbalanced motions can overlap strongly in wavenumber space; a fixed low-pass or Gaussian filter cannot reliably separate them when their scales are not well separated. Nevertheless, the target of such a model is the balanced component of SSH, not the full instantaneous velocity field. If the desired output is total velocity, then removing unbalanced SSH may improve balanced velocity estimates while also discarding wave-related velocity signals. Conversely, if the desired output is only the balanced velocity, the problem becomes better posed but remains dependent on how ``balanced'' and ``unbalanced'' components are defined in the training data.  In most cases, the balanced and unbalanced components of an SSH field cannot be separated cleanly at submesoscales even with high-frequency hourly sampling. This is the reason we choose to directly predict velocity vector instead of separating balanced and unbalanced SSH, which we also tried but not reported in this study. 

More recently, \citet{Zhou2025} addressed the problem using actual SWOT observations and collocated drifter data to infer sea-surface velocity. This study is an important step because it moves beyond idealized or model-only experiments and directly uses the observational geometry and error characteristics of SWOT. By learning from drifter-constrained velocities, such an approach can potentially correct systematic departures from geostrophy, including those associated with low latitudes or unresolved ageostrophic contributions. However, drifter-supervised learning introduces a different set of assumptions. Drifters sample particular flow regimes and geographical regions unevenly, may include wind-driven and wave-related motions depending on the drifter type and processing, and provide velocities at finite depth and finite temporal resolution rather than an instantaneous Eulerian surface velocity exactly collocated with the SSH measurement. Therefore, success against drifter validation demonstrates practical predictive skill, but it does not by itself prove that SSH alone uniquely determines the target velocity. The global statistics from \citet{Zhou2025} serve as a first demonstration but more regional scrutiny is needed for generality.

Taken together, these studies show that neural networks can learn nonlocal spatial derivatives of SSH, empirical relationships between wind forcing and currents, and statistical priors that are difficult to express as analytic formulas, and that they can improve on geostrophic estimates. Three aspects of this work motivate the present study. First, in submesoscale-permitting simulations, previous studies reconstruct quantities derived from the velocity, such as vorticity, strain, and divergence \citep{Xiao2023}, or the balanced part of SSH \citep{Gao2024,Lyu2024,Wang2026}, rather than the velocity vector itself. We predict the velocity vector directly, both at 53\,m depth and at the surface, and derive vorticity, divergence, and strain from it. Because geostrophy already explains most of the velocity variance at 53\,m, the network predicts the departure from geostrophy and only has to learn what geostrophy misses (Section~\ref{sec:residual}). Second, the studies differ in their inputs, targets, and regions at the same time: some use SSH alone \citep{Xiao2023,Gao2024,Lyu2024}, whereas others add SST \citep{Martin2024} or SST, sea-surface salinity, wind stress, and geographical coordinates \citep{Sinha2021}. How much each input contributes to the velocity has therefore not been compared in a single submesoscale-permitting setting. With one region and one network, we compare raw SSH with SSH gradients and add SST and wind stress one at a time, at 53\,m and at the surface (Sections~\ref{sec:ablation} and \ref{sec:surface}). Third, studies that predict velocity evaluate it mainly by pointwise errors \citep{Sinha2021,Zhou2025}, so it is not known whether the predicted velocity has the correct spatial derivatives and transfers kinetic energy across scales as the true flow does. Likewise, altimetric estimates of this cross-scale transfer use geostrophic velocities \citep{Scott2005}, whose error in the transfer can be measured only where the true velocity is known. We evaluate the vorticity and divergence, the joint distribution of vorticity and strain, and the spectral kinetic-energy flux of the predicted and the geostrophic velocities against the LLC4320 velocity on a period not used for training (Sections~\ref{sec:kinematic_results}, \ref{sec:validation_results}, and \ref{sec:flux}).

\section{Methods and Data}
\label{sec:architecture}

\subsection{Data}
\label{sec:data}

\begin{figure}[t]
\centering
\includegraphics[width=\textwidth]{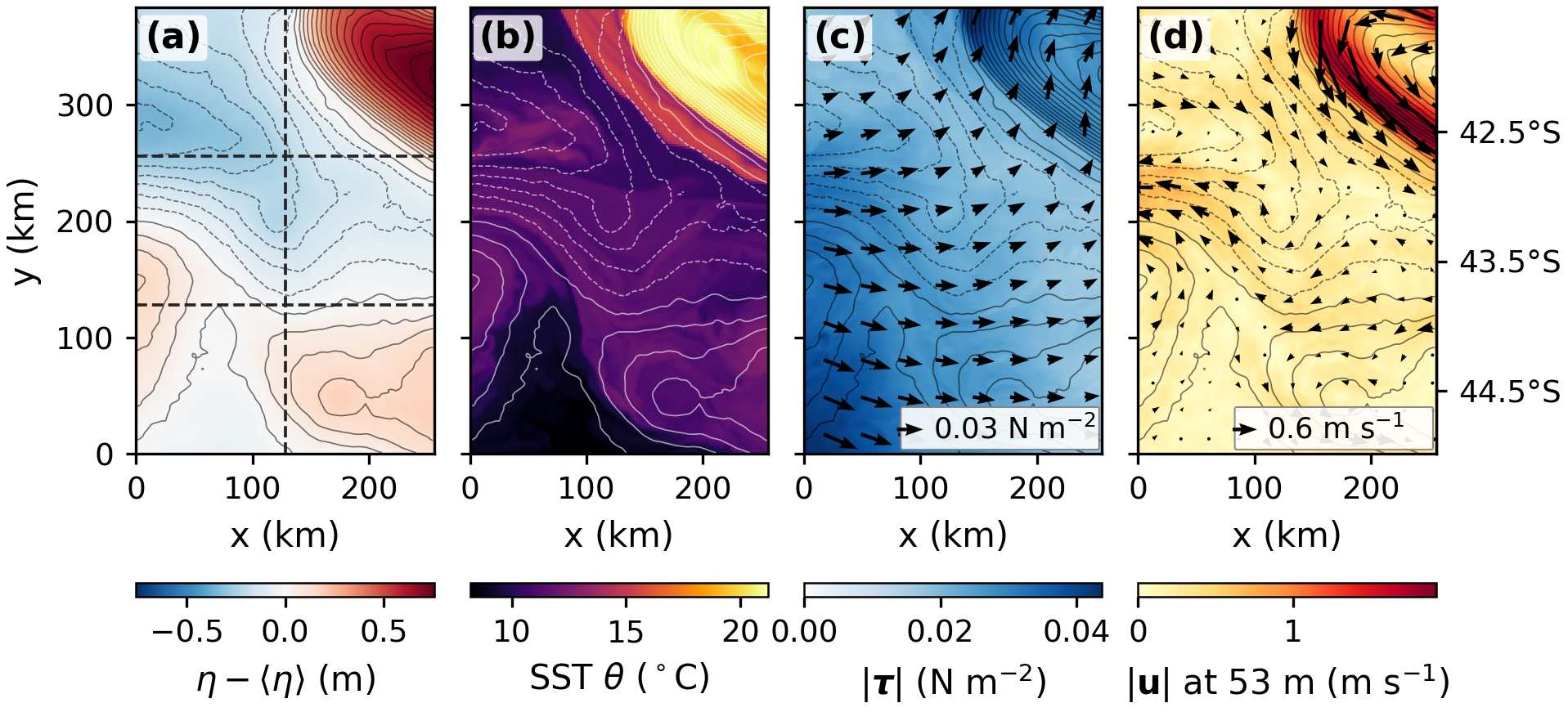}
\caption{Snapshot of the fields used in this study at 10:00 UTC on 10 April 2012, on the 2\,km analysis grid ($x$ eastward from 11.8$^\circ$E, $y$ northward from 45.0$^\circ$S; latitude on the right). (a)~Sea-surface height anomaly minus its domain mean; (b)~SST; (c)~wind stress (vectors, every 16th grid point) over its magnitude; (d)~velocity at 53\,m depth (vectors, every 12th grid point) over speed. Contours show $\eta$ every 0.05\,m on all panels (dashed where below the domain mean). Dashed lines in (a) mark the six fixed $128\times128$\,km subdomains; validation and test patches are these tiles, whereas training samples are drawn at random positions in the domain (Section~\ref{sec:augmentation}). This snapshot is from the training period.}
\label{fig:snapshot}
\end{figure}

We use the pre-SWOT regional extraction of the LLC4320 simulation for the ``Cape Basin'' region (PO.DAAC product \nolinkurl{MITgcm_LLC4320_Pre-SWOT_JPL_L4_CapeBasin_v1.0}), which covers 11.8--15.8$^\circ$E and 41.0--45.0$^\circ$S. LLC4320 is a global MITgcm simulation with $1/48^{\circ}$ nominal horizontal resolution and 90 vertical levels, forced by 6-hourly ERA-Interim atmospheric reanalysis and by a tidal forcing representing the 16 largest tidal constituents \citep{Rocha2016}. The Cape Basin subset used in this study can be downloaded from \url{https://doi.org/10.5067/PRESW-CBJ10}. The region lies southwest of the Agulhas retroflection and contains energetic eddies and sharp SST fronts (Fig.~\ref{fig:snapshot}); in this region the model grid is aligned with longitude and latitude and its spacing is about 1.6--1.7\,km in the zonal and 1.6\,km in the meridional direction. The regional product provides hourly fields from 13 September 2011 to 15 November 2012 (429 daily files). From each hourly snapshot we use the sea-surface height anomaly $\eta$ (model variable \texttt{Eta}), the potential temperature of the top model level (0.5\,m, hereafter SST, $\theta$), the ocean surface stress $(\tau_x,\tau_y)$ (\texttt{oceTAUX}, \texttt{oceTAUY}; hereafter wind stress), and the horizontal velocity at model level~16, whose cell center is at 52.7\,m depth (hereafter $(u,v)$, with $u$ eastward and $v$ northward). For the experiments of Section~\ref{sec:surface}, the target is instead the velocity of the top model level (0.5\,m), hereafter the surface velocity.

Each snapshot is interpolated linearly, using a Delaunay triangulation of the model grid points, onto a regular 2\,km Cartesian grid whose coordinates are the cumulative model grid spacings; the staggering of velocity and stress on the model's C-grid (half a model cell, $\approx$0.8\,km) is not accounted for. We tile the regridded domain into non-overlapping $64\times64$ patches ($128\times128$\,km) and keep only patches that are completely covered by data. This leaves six fixed subdomains, arranged three (south to north) by two (west to east), which together cover 256\,km\,$\times$\,384\,km (Fig.~\ref{fig:snapshot}). After discarding the first and last 50 hours, 10{,}196 hourly time steps remain, giving 61{,}176 patches. The region contains no land, so every pixel is valid.

The inputs and the target have very different frequency content. Because LLC4320 is tidally forced, $\eta$ at a fixed point is dominated by the semidiurnal tide (88\,\% of its variance at periods of 11.5--13\,h, 11\,\% at periods longer than two days). The 53\,m velocity, in contrast, is dominated by slowly varying flow: 92\,\% of its variance is at periods longer than two days, 2\,\% in the semidiurnal band and 1.5\,\% in the near-inertial band (inertial period 17.5\,h). The derived fields differ again: 64\,\% of the variance of $\zeta/f$ at 53\,m is at periods longer than two days, whereas 63\,\% of the variance of the divergence $\delta/f$ is at periods shorter than 13\,h, i.e., in internal tides and other internal gravity waves. (These fractions are from frequency spectra at 12 fixed points over the full record.) Speeds at 53\,m have a median of 0.22\,m/s and a 99th percentile of 0.86\,m/s; SST ranges from about 6 to 18$^\circ$C (1st--99th percentiles).

\subsection{Training, Validation, and Test Periods}
\label{sec:split}

Hourly fields change slowly (the RMSE of speed between 53\,m velocity fields one hour apart is 0.024\,m/s), so a random split of patches would place nearly identical samples in training and evaluation and overstate skill about fivefold (Appendix~\ref{sec:leakage}). We therefore split the record into three consecutive blocks of hours, separated by 48-hour gaps so that no training hour lies within two days of a validation or test hour (Table~\ref{tab:split}). The training block contains about the first 70\,\% of the record, the validation block the next 10\,\%, and the test block the remaining 20\,\% (Table~\ref{tab:split}). The validation block is used for early stopping and for choosing between configurations; each trained model is evaluated on the test block once, with the checkpoint selected on the validation block. The final model is chosen by validation score (Section~\ref{sec:generalization}). The test period (late August to mid-November 2012) covers the same season as the first two months of the training block (mid-September to November 2011), one year later.

\begin{table}[h]
\centering
\caption{Temporal split of the 10{,}196 hourly time steps. Each hour contributes six $64\times64$ patches.}
\label{tab:split}
\begin{tabular}{lccc}
\toprule
Block & Period & Hours & Patches \\
\midrule
Training & 15 September 2011 -- 8 July 2012 & 7{,}137 & 42{,}822 \\
\multicolumn{4}{c}{\footnotesize 48-hour gap} \\
Validation & 10 July -- 19 August 2012 & 971 & 5{,}826 \\
\multicolumn{4}{c}{\footnotesize 48-hour gap} \\
Test & 21 August -- 12 November 2012 & 1{,}992 & 11{,}952 \\
\bottomrule
\end{tabular}
\end{table}

\paragraph{Normalization.} Each input and target channel is standardized with a mean and standard deviation estimated from 4{,}096 randomly drawn training patches (after augmentation, when rotations are used; Section~\ref{sec:augmentation}). No validation or test samples enter these statistics.

\subsection{Problem Formulation}

We formulate SSH-to-velocity inference as a supervised dense prediction problem on fixed-size patches. Each sample is a $64\times 64$ grid with $C_{in}$ input channels and two output channels $(u,v)$. Let
\begin{equation}
	\mathbf{x}\in\mathbb{R}^{C_{in}\times H\times W},
	\qquad
	\mathbf{y}\in\mathbb{R}^{2\times H\times W},
\end{equation}
with $H=W=64$. The network learns a mapping
\begin{equation}
	f_{\theta}:\mathbf{x}\mapsto\hat{\mathbf{y}},
\end{equation}
where $\hat{\mathbf{y}}=(\hat{u},\hat{v})$ is the predicted horizontal velocity field.

\subsection{Inputs}

We test two representations of SSH, each combined with optional covariates:
\begin{equation}
	\mathbf{x}_{\eta}\in\{\eta,\;\eta+\theta,\;\eta+\boldsymbol{\tau},\;\eta+\theta+\boldsymbol{\tau}\},
	\qquad
	\mathbf{x}_{\nabla\eta}\in\{\nabla\eta,\;\nabla\eta+\theta,\;\nabla\eta+\boldsymbol{\tau},\;\nabla\eta+\theta+\boldsymbol{\tau}\},
\end{equation}
where $\eta$ is the SSH anomaly, $\nabla\eta=(\partial\eta/\partial x,\,\partial\eta/\partial y)$ its horizontal gradient, $\theta$ is SST, and $\boldsymbol{\tau}=(\tau_x,\tau_y)$ is the wind stress. The gradients are computed with centered differences on the 2\,km grid, with replicated values beyond the patch edges. All inputs are computed from the (augmented) raw fields of each patch during training.

\subsection{Target: Velocity or Departure from Geostrophy}
\label{sec:residual}

The network either predicts the velocity directly or predicts its departure from the $f$-plane geostrophic velocity computed from the same SSH,
\begin{equation}
	\mathbf{r} = \mathbf{u} - \mathbf{u}_g,
	\qquad
	\mathbf{u}_g = \frac{g}{f_0}\,\hat{\mathbf{z}}\times\nabla\eta,
	\label{eq:residual}
\end{equation}
in which case the predicted velocity is $\hat{\mathbf{u}}=\mathbf{u}_g+\hat{\mathbf{r}}$. Here $f_0=2\Omega\sin(-43^\circ)=-9.95\times10^{-5}$\,s$^{-1}$ is the Coriolis parameter at the mean latitude of the domain ($f$ varies by less than $\pm4$\,\% across the domain). Geostrophy already explains most of the velocity variance at 53\,m (Section~\ref{sec:kinematic_results}), so the network only has to learn the remainder that is approximately the ageostrophic component.

\subsection{Data Augmentation}
\label{sec:augmentation}

The training data contain only six fixed subdomains, so the network sees relatively few independent flow configurations. We use two label-preserving augmentations.

\paragraph{Rotations.} Each training patch is rotated counterclockwise by a random multiple of $90^\circ$. Scalar fields ($\eta$, $\theta$) are rotated as images; for vector fields ($\boldsymbol{\tau}$, $\mathbf{u}$), the components are rotated as well, $(u,v)\rightarrow(-v,u)$ for each $90^\circ$ step. On an $f$-plane the equations of motion are invariant under such rotations, and we verified that the rotated velocity remains exactly the geostrophic velocity of the rotated SSH. Reflections are not used: a mirror image reverses the sense of rotation and is equivalent to changing the sign of $f$.

\paragraph{Random crops.} Instead of the six fixed tiles, each training sample is a $64\times64$ window at a random position within the $192\times128$-cell domain of a random training hour. This exposes the network to many more distinct windows of the same fields. Validation and test patches are always the six fixed tiles.

\subsection{Network Backbone}

\begin{figure}[t]
\centering
\includegraphics[width=\textwidth]{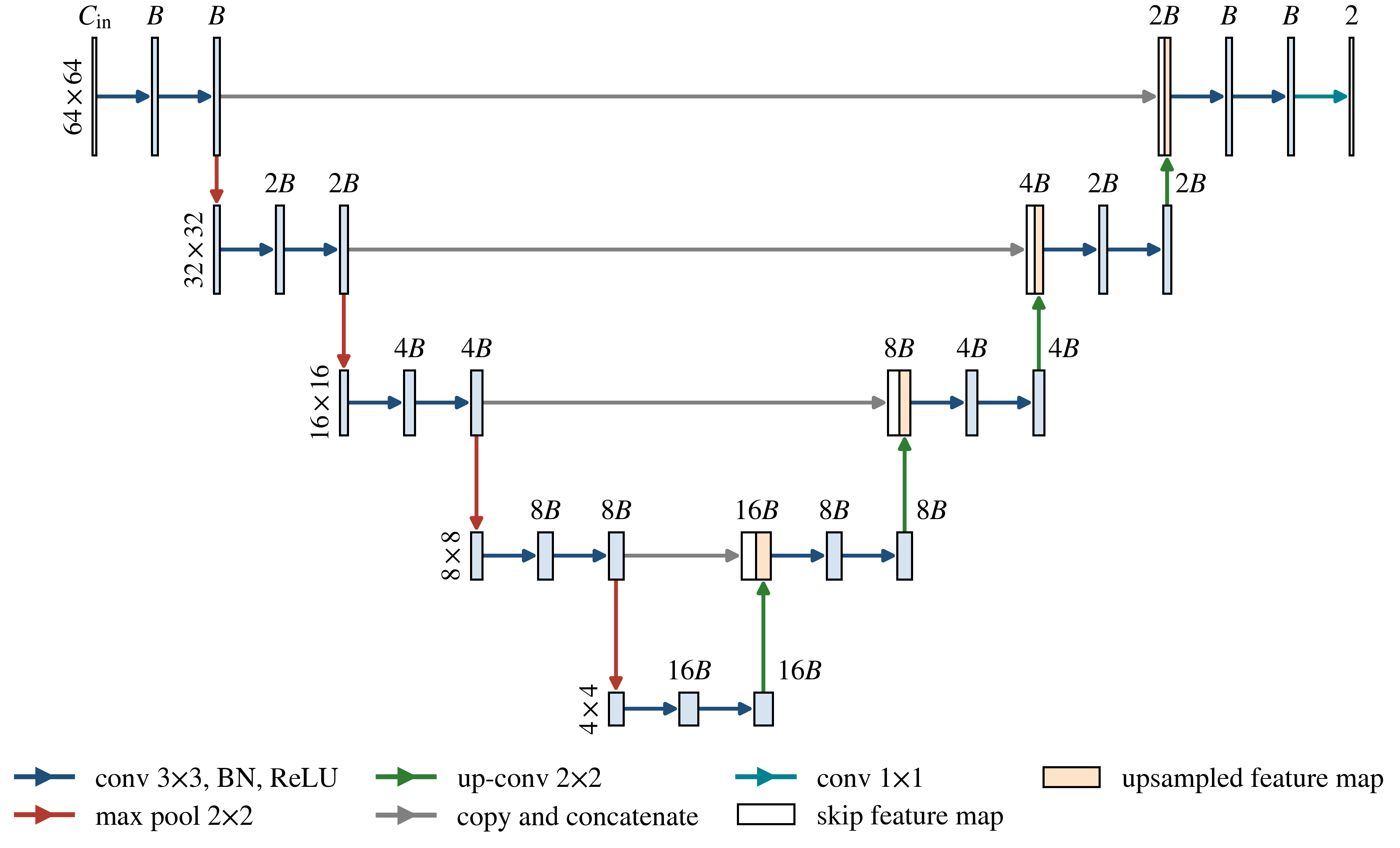}
\caption{U-Net encoder--decoder architecture. Blue: encoder stages; green: decoder stages; orange: bottleneck. Each block applies two ($3\times3$ convolution, batch normalization, ReLU) layers; labels give the number of channels and the spatial size. Encoder stages are followed by $2\times2$ max-pooling, decoder stages are preceded by a $2\times2$ transposed convolution, and dashed arrows are skip connections (copy and concatenate). $C_\mathrm{in}$ is the number of input channels (1--5) and $B\in\{64,128,192\}$ the base width.}
\label{fig:unet_architecture}
\end{figure}

We use a U-Net encoder--decoder \citep{Ronneberger2015} with skip connections (Fig.~\ref{fig:unet_architecture}). Each block consists of two $3\times 3$ convolutions, each followed by batch normalization and a ReLU. Four encoder stages, each followed by $2\times2$ max-pooling, lead to a bottleneck at $4\times4$; four decoder stages upsample with $2\times2$ transposed convolutions and concatenate the encoder features of the same resolution. A final $1\times 1$ convolution maps to the two output channels.

Model capacity is controlled by the base channel width $B$, with channel widths
\begin{equation}
	[B,2B,4B,8B,16B]
\end{equation}
from the first encoder stage to the bottleneck. We compare $B{=}64$ ($3.1\times10^7$ parameters), $B{=}128$ ($1.2\times10^8$), and $B{=}192$ ($2.8\times10^8$).

\subsection{Loss Functions}

The base loss is the mean squared error (MSE) or mean absolute error (MAE) between the standardized prediction and target, averaged over all pixels and both components. Optionally, we add penalties on the errors of relative vorticity and horizontal divergence of the predicted velocity,
\begin{equation}
	\mathcal{L}=\mathcal{L}_{\text{base}}
	+ w_\zeta\,\mathcal{L}_{\text{base}}\!\left(\frac{\hat\zeta}{f_0 s_\zeta},\frac{\zeta}{f_0 s_\zeta}\right)
	+ w_\delta\,\mathcal{L}_{\text{base}}\!\left(\frac{\hat\delta}{f_0 s_\delta},\frac{\delta}{f_0 s_\delta}\right),
	\label{eq:loss}
\end{equation}
where $\zeta$ and $\delta$ are defined in Eq.~\eqref{eq:vorticity} and computed from the velocity in physical units, $s_\zeta$ and $s_\delta$ are the standard deviations of $\zeta/f_0$ and $\delta/f_0$ in the training sample used for normalization, so that both penalty terms are of order one, and $w_\zeta=w_\delta=0.5$ unless stated otherwise.

\subsection{Optimization and Training Protocol}

All models are trained with AdamW \citep{Loshchilov2019} (learning rate $3\times10^{-4}$, weight decay $10^{-4}$, batch size 128), with a linear learning-rate warmup over the first two epochs followed by cosine decay over the maximum number of epochs (60, or 100 for the longer runs), gradient-norm clipping at 1.0, and bfloat16 mixed precision. One epoch is 334 batches, equivalent to one pass over the training tiles. After each epoch we compute the RMSE of speed on the validation block; training stops when it has not improved for 10 epochs (20 for the longer runs), and the checkpoint with the lowest value is kept. Every configuration is trained with three random seeds, and we report the mean and sample standard deviation over seeds; the only exception is the random-split comparison in Appendix~\ref{sec:leakage}, which was trained once.

\subsection{Evaluation Metrics}

We report the RMSE of each velocity component, RMSE$_u$ and RMSE$_v$, over the test block. Our primary metric is the RMSE of speed,
\begin{equation}
	\text{RMSE speed} = \sqrt{\frac{1}{N}\sum\left(\sqrt{\hat{u}^2+\hat{v}^2} - \sqrt{u^2+v^2}\right)^2},
\end{equation}
where the sum runs over all $N$ test pixels. We also report the direction error, the mean absolute angle between predicted and target velocity vectors over pixels with target speed above 0.01\,m/s.

\subsection{Kinematic Evaluation Metrics}
\label{sec:kinematic_metrics}

Beyond pointwise velocity errors, we evaluate the relative vorticity and horizontal divergence of the predicted fields, computed with centered differences on the 2\,km grid (with replicated values beyond the patch edges):
\begin{equation}
	\zeta = \frac{\partial v}{\partial x} - \frac{\partial u}{\partial y},
	\qquad
	\delta = \nabla\cdot\mathbf{u} = \frac{\partial u}{\partial x} + \frac{\partial v}{\partial y}.
	\label{eq:vorticity}
\end{equation}
Vorticity characterizes the rotational part of the flow; under geostrophic balance $\zeta_g = (g/f)\,\nabla^2\eta$, so it is directly constrained by SSH curvature. Divergence is ageostrophic: the $f$-plane geostrophic flow is non-divergent, so any divergence in a prediction reflects learned ageostrophic structure.

\paragraph{Normalization and sign convention.} We normalize both quantities by $f_0$ and keep its sign: because $f_0<0$, $\zeta/f>0$ denotes cyclonic and $\zeta/f<0$ anticyclonic rotation, so the sign of $\zeta/f$ has the same meaning in both hemispheres. Non-negative quantities such as the strain rate are normalized by $|f|$. We report the RMSE of $\zeta/f$ and $\delta/f$ and the Pearson correlation $r$ between predicted and true fields over all test pixels.

\paragraph{Geostrophic baseline.}
As a physical reference we compute the $f$-plane geostrophic velocity $\mathbf{u}_g$ of Eq.~\eqref{eq:residual} from $\eta$ on the same patches. This is the geostrophic velocity at the surface; the geostrophic velocity at 53\,m differs from it by the thermal-wind shear over the upper 53\,m, which the baseline ignores. Its divergence is zero, so its RMSE of $\delta/f$ equals the standard deviation of the true $\delta/f$.

\subsection{Experiments}
\label{sec:experiments}

We proceed in two steps. First (Section~\ref{sec:generalization}), starting from a baseline U-Net ($B{=}64$; SSH gradients, SST, and wind stress as inputs; velocity as target; MSE loss; no augmentation), we add one training choice at a time and keep those that improve the validation error. Second (Section~\ref{sec:ablation}), we repeat the input, capacity, and loss ablations around the final configuration. Appendix~\ref{app:all_runs} lists all experiments.

\section{Results}
\label{sec:training_results}

\subsection{Toward a Model That Generalizes}
\label{sec:generalization}

\begin{figure}[t]
\centering
\includegraphics[width=\textwidth]{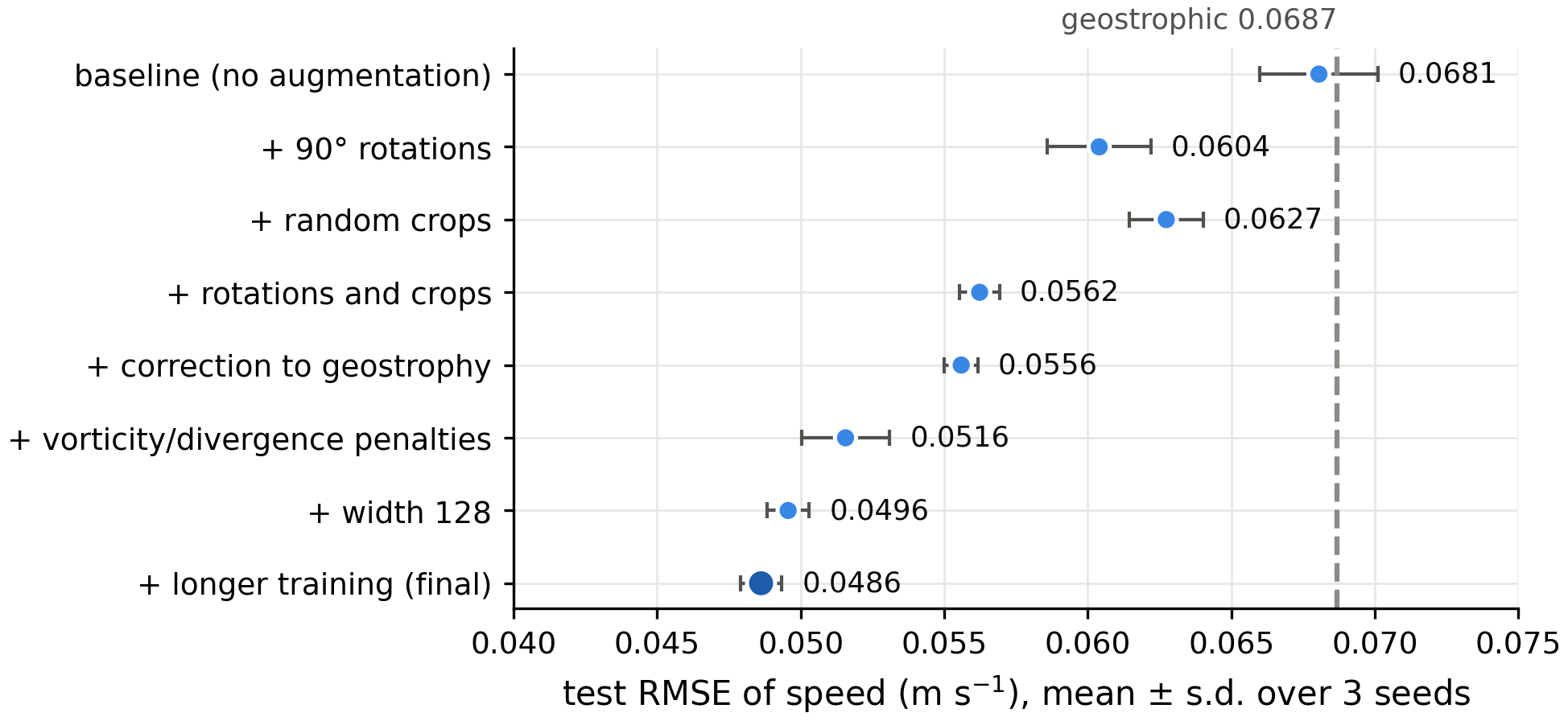}
\caption{Test RMSE of speed (mean and standard deviation over three seeds) as training choices are added (Table~\ref{tab:progression}). The dashed line is the $f$-plane geostrophic estimate.}
\label{fig:progression}
\end{figure}

Table~\ref{tab:progression} and Figure~\ref{fig:progression} show how the test error changes as training choices are added. The baseline U-Net is no better than geostrophy (0.0681 versus 0.0687\,m/s), and its validation error is lowest after one to five epochs while the training error keeps decreasing: it starts to memorize the training period almost immediately.

\begin{table}[t]
\centering
\caption{Test and validation RMSE of speed (m/s; mean $\pm$ standard deviation over three seeds), test direction error, and test correlations with the true vorticity and divergence as training choices are added. All U-Nets use SSH gradients, SST, and wind stress. E1--E3 add the augmentations to E0 separately and together; from E4 on, each row adds one choice to the row above it.}
\label{tab:progression}
\setlength{\tabcolsep}{4pt}
\begin{tabular}{llccccc}
\toprule
 & Configuration & Test RMSE speed & Val.\ RMSE speed & Dir.\ error & $r_\zeta$ & $r_\delta$ \\
\midrule
 & Geostrophic estimate & 0.0687 & 0.0768 & 18.4$^\circ$ & 0.749 & 0 \\
\midrule
E0 & Baseline ($B{=}64$, no augmentation) & 0.0681 $\pm$ 0.0021 & 0.0740 & 19.0$^\circ$ & 0.756 & 0.109 \\
E1 & E0 + rotations & 0.0604 $\pm$ 0.0018 & 0.0632 & 16.9$^\circ$ & 0.812 & 0.169 \\
E2 & E0 + random crops & 0.0627 $\pm$ 0.0013 & 0.0660 & 17.5$^\circ$ & 0.791 & 0.167 \\
E3 & E0 + rotations + crops & 0.0562 $\pm$ 0.0007 & 0.0577 & 15.9$^\circ$ & 0.861 & 0.302 \\
E4 & E3, departure from geostrophy & 0.0556 $\pm$ 0.0006 & 0.0545 & 15.1$^\circ$ & 0.896 & 0.394 \\
R3b & E4 + vorticity/divergence penalties & 0.0516 $\pm$ 0.0015 & 0.0513 & 14.4$^\circ$ & 0.921 & 0.627 \\
C1 & R3b with $B{=}128$ & 0.0496 $\pm$ 0.0007 & 0.0496 & 14.0$^\circ$ & 0.930 & 0.661 \\
C3 & C1, longer training (final model) & \textbf{0.0486 $\pm$ 0.0007} & \textbf{0.0490} & 14.0$^\circ$ & 0.931 & 0.661 \\
\bottomrule
\end{tabular}
\end{table}

\paragraph{Augmentation.} Rotations reduce the test error by 11\,\% and random crops by 8\,\%; together they reduce it by 17\,\%, to 0.0562\,m/s. With both augmentations, the validation error keeps improving for about 20 epochs instead of a few, so augmentation largely removes the early memorization.

\paragraph{Departure from geostrophy.} Predicting $\mathbf{u}-\mathbf{u}_g$ instead of $\mathbf{u}$ reduces the RMSE of speed only slightly (0.0556 versus 0.0562\,m/s) but improves the derived fields clearly: $r_\zeta$ increases from 0.861 to 0.896 and $r_\delta$ from 0.302 to 0.394. Supplying $\mathbf{u}_g$ as two additional input channels instead of using it in the target helps less (0.0573\,m/s; Appendix~\ref{app:all_runs}).

\paragraph{Vorticity and divergence penalties.} Adding the penalties of Eq.~\eqref{eq:loss} reduces the RMSE of speed by a further 7\,\% (to 0.0516\,m/s) and raises $r_\delta$ from 0.394 to 0.627, the largest single improvement in divergence.

\paragraph{Capacity and training length.} Increasing the base width from 64 to 128 reduces the error by another 4\,\%; a width of 192 performs the same as 128 (Section~\ref{sec:ablation}). Training for up to 100 epochs with a patience of 20 gives a further small gain (0.0486\,m/s). Doubling the penalty weights, smoothing the SSH before computing $\mathbf{u}_g$ (Gaussian of 4 or 8\,km), or using an MAE base loss did not help (Appendix~\ref{app:all_runs}).

\paragraph{Model selection.} Among the round-3 candidates (Appendix~\ref{app:all_runs}), the configuration with $B{=}128$ and longer training (C3) and the configuration with $B{=}192$ and standard training (C2) have the same mean validation error (0.0490\,m/s); we select C3, the smaller of the two. In the later capacity ablation, C3 trained at $B{=}192$ reached a mean validation error of 0.0486\,m/s (Table~\ref{tab:capacity}); this difference is smaller than the spread over seeds (0.0478--0.0492\,m/s), and its test error (0.0490\,m/s) is not lower than that of C3. Its test RMSE of speed is $0.0486\pm0.0007$\,m/s, with RMSE$_u=0.0460\pm0.0003$\,m/s and RMSE$_v=0.0464\pm0.0003$\,m/s, compared with 0.0645 and 0.0655\,m/s for geostrophy. The velocity components in the test block have standard deviations of 0.20 and 0.22\,m/s, so the final model explains about 95\,\% of their variance ($R^2_u=0.947$, $R^2_v=0.954$), compared with 90\,\% for geostrophy.

\subsection{Ablations Around the Final Model}
\label{sec:ablation}

Starting from the final configuration (C3), we change one factor at a time and keep all other settings, including the longer training (up to 100 epochs, patience 20), fixed (Tables~\ref{tab:ablation}--\ref{tab:loss_ablation}). These ablation runs are therefore not the same runs as C1 and C2 in Appendix~\ref{app:all_runs}, which use standard training (up to 60 epochs, patience 10).

\begin{table}[t]
\centering
\caption{Test RMSE of speed (m/s; mean $\pm$ standard deviation over three seeds) for raw SSH and SSH gradients combined with each covariate set ($B{=}128$, departure from geostrophy, augmentation, MSE with vorticity/divergence penalties, longer training as in C3; the gradient model with SST and wind stress is C3 itself). The change is that of the gradient model relative to the raw-SSH model. Validation RMSE of speed in parentheses; test correlations $r_\zeta$ / $r_\delta$ below.}
\label{tab:ablation}
\setlength{\tabcolsep}{5pt}
\begin{tabular}{lccc}
\toprule
Covariates & Raw SSH $\eta$ & SSH gradients $\nabla\eta$ & Change \\
\midrule
none & 0.0597 $\pm$ 0.0002 (0.0650) & 0.0551 $\pm$ 0.0015 (0.0606) & $-$7.6\,\% \\
 & 0.801 / 0.249 & 0.868 / 0.417 & \\
SST & 0.0592 $\pm$ 0.0018 (0.0597) & 0.0581 $\pm$ 0.0020 (0.0562) & $-$1.9\,\% \\
 & 0.830 / 0.334 & 0.881 / 0.478 & \\
wind stress & 0.0487 $\pm$ 0.0005 (0.0527) & 0.0489 $\pm$ 0.0005 (0.0521) & $+$0.3\,\% \\
 & 0.892 / 0.659 & 0.927 / 0.644 & \\
SST + wind stress & 0.0510 $\pm$ 0.0002 (0.0508) & \textbf{0.0486 $\pm$ 0.0007 (0.0490)} & $-$4.6\,\% \\
 & 0.901 / 0.684 & 0.931 / 0.661 & \\
\bottomrule
\end{tabular}
\end{table}

\paragraph{Input representation.} SSH gradients give a clearly lower RMSE of speed than raw SSH without covariates ($-$7.6\,\%) and with SST and wind stress ($-$4.6\,\%). With SST alone ($-$1.9\,\%) and with wind stress alone ($+$0.3\,\%), the difference is smaller than the spread over seeds, so the two representations perform equally there. In contrast, gradients give a higher vorticity correlation for every covariate set ($r_\zeta$ = 0.87--0.93 versus 0.80--0.90), Their divergence correlation is higher without wind stress (0.42 versus 0.25 and 0.48 versus 0.33) but slightly lower with it (0.64 versus 0.66 and 0.66 versus 0.68). A plausible reason is that geostrophic velocity is set by the SSH gradient rather than by the absolute sea level, and that most of the temporal variability of $\eta$ is the semidiurnal tide (Section~\ref{sec:data}), which in the open ocean is mainly barotropic, with a length scale much larger than a patch; within a patch it therefore appears mainly as a nearly uniform offset that the gradients remove.

\paragraph{Covariates.} Wind stress is the most useful covariate. Added to SSH gradients alone it reduces the test RMSE of speed by 11\,\% (from 0.0551 to 0.0489\,m/s) and raises $r_\delta$ from 0.42 to 0.64; added to raw SSH it reduces the error by 18\,\%. Wind stress drives Ekman and near-inertial currents in the surface boundary layer; in the model, the KPP boundary-layer depth exceeds 53\,m at most of the sampled times in austral winter (May--September) but at fewer than about a third of them in summer (December--March; one day sampled per month), so these currents reach the target depth mainly in winter, and near-inertial motions account for only 1.5\,\% of the velocity variance at 53\,m. The effect of SST depends on the evaluation period. On the validation block (July--August 2012), adding SST reduces the RMSE of speed in all four comparisons of Table~\ref{tab:ablation} (for example from 0.0521 to 0.0490\,m/s with gradients and wind stress). On the test block (August--November 2012), its effect is mixed: it changes the error by less than the spread over seeds with gradients and wind stress (0.0489 to 0.0486\,m/s) and with raw SSH alone (0.0597 to 0.0592\,m/s), and it increases the error with gradients alone (0.0551 to 0.0581\,m/s) and with raw SSH and wind stress (0.0487 to 0.0510\,m/s). The same pattern appears in the earlier rounds of experiments: without SST, the validation error is higher but the test error lower (R2a versus E4, C5 versus C1; Appendix~\ref{app:all_runs}). Physically, SST marks surface density fronts, and through thermal-wind balance horizontal density gradients set the vertical shear between the surface geostrophic velocity implied by $\eta$ and the velocity at 53\,m; how useful this information is may depend on the seasonal stratification. We keep SST in the final model because the model was selected on the validation block; without SST, the corresponding configuration has a test RMSE of speed of 0.0489\,m/s, statistically indistinguishable from the final model.

\begin{table}[h]
\centering
\caption{Capacity: metrics (mean $\pm$ standard deviation over three seeds) for the final configuration at three base widths, all with longer training (up to 100 epochs, patience 20). The $B{=}128$ row is C3; the $B{=}64$ and $B{=}192$ rows are separate runs and differ from R3b and C2 in Appendix~\ref{app:all_runs}, which use standard training.}
\label{tab:capacity}
\setlength{\tabcolsep}{5pt}
\begin{tabular}{lccccc}
\toprule
Base width (parameters) & Test RMSE speed & Val.\ RMSE speed & Dir.\ error & $r_\zeta$ & $r_\delta$ \\
\midrule
64 ($3.1\times10^7$) & 0.0511 $\pm$ 0.0009 & 0.0512 & 14.4$^\circ$ & 0.921 & 0.614 \\
128 ($1.2\times10^8$, final) & \textbf{0.0486 $\pm$ 0.0007} & 0.0490 & 14.0$^\circ$ & 0.931 & 0.661 \\
192 ($2.8\times10^8$) & 0.0490 $\pm$ 0.0020 & 0.0486 & 13.9$^\circ$ & 0.936 & 0.686 \\
\bottomrule
\end{tabular}
\end{table}

\paragraph{Capacity.} Increasing the base width from 64 to 128 reduces the test RMSE of speed by 5\,\%; a width of 192 gives no further improvement in RMSE of speed and only slightly higher correlations (Table~\ref{tab:capacity}).

\begin{table}[h]
\centering
\caption{Training objective: metrics (mean $\pm$ standard deviation over three seeds) for the final configuration with four losses, all with longer training (up to 100 epochs, patience 20). The last row is C3.}
\label{tab:loss_ablation}
\setlength{\tabcolsep}{5pt}
\begin{tabular}{lccccc}
\toprule
Loss & Test RMSE speed & Val.\ RMSE speed & Dir.\ error & $r_\zeta$ & $r_\delta$ \\
\midrule
MSE & 0.0533 $\pm$ 0.0009 & 0.0520 & 14.4$^\circ$ & 0.918 & 0.545 \\
MAE & 0.0526 $\pm$ 0.0009 & 0.0520 & 14.4$^\circ$ & 0.920 & 0.562 \\
MAE + penalties & 0.0512 $\pm$ 0.0007 & 0.0494 & 14.2$^\circ$ & 0.934 & 0.674 \\
MSE + penalties (final) & \textbf{0.0486 $\pm$ 0.0007} & 0.0490 & 14.0$^\circ$ & 0.931 & 0.661 \\
\bottomrule
\end{tabular}
\end{table}

\paragraph{Training objective.} MSE and MAE perform equally well (0.0533 and 0.0526\,m/s, a difference within the spread over seeds). The vorticity and divergence penalties reduce the RMSE of speed by 9\,\% with MSE and 3\,\% with MAE, and raise $r_\delta$ from 0.55 to 0.66 (Table~\ref{tab:loss_ablation}).

\subsection{Example Predictions}

\begin{figure}[t]
\centering
\includegraphics[width=\textwidth]{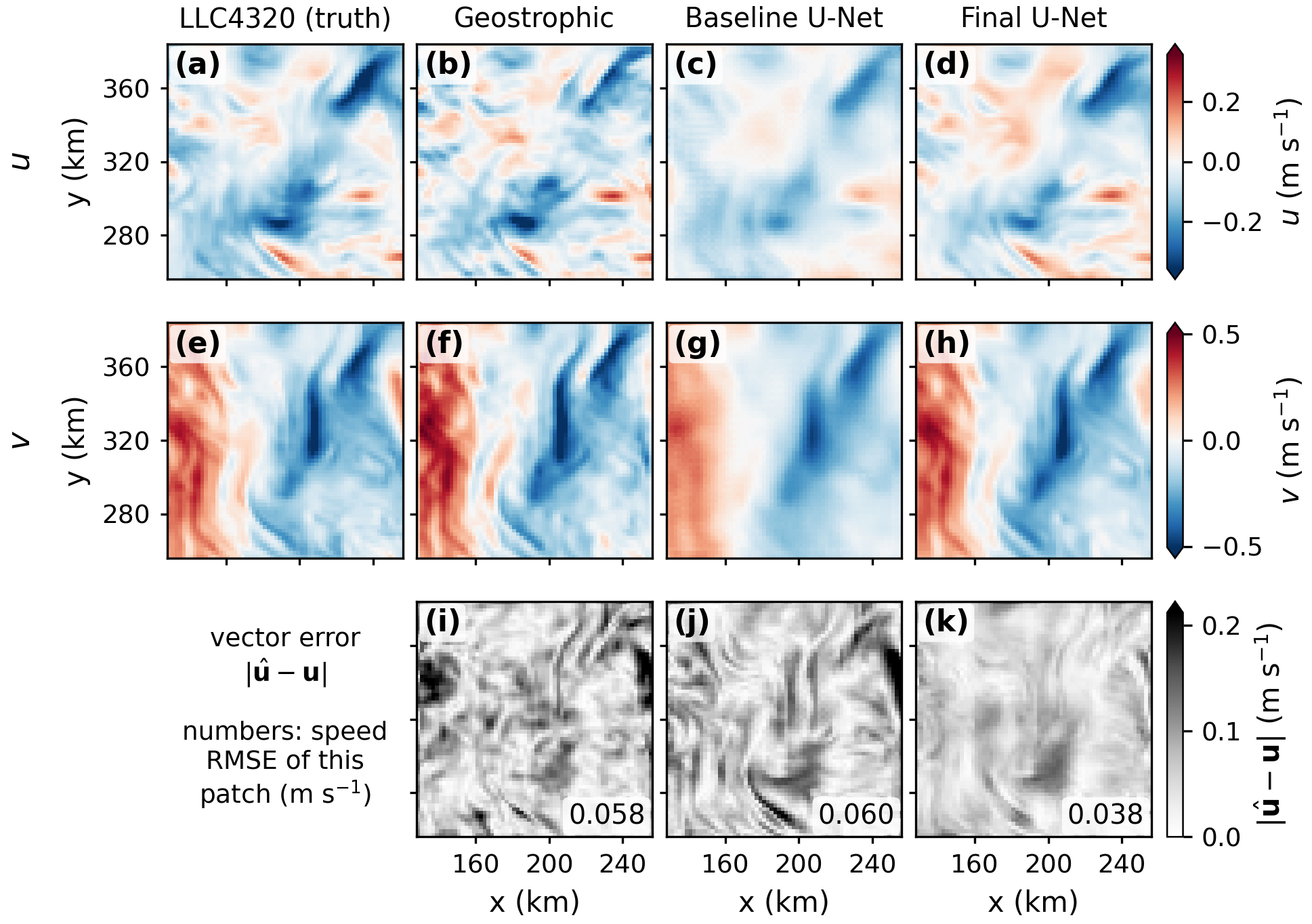}
\caption{Velocity on a representative test patch ($128\times128$\,km; the same patch as in Fig.~\ref{fig:kinematic_diagnostics}). Columns: LLC4320 truth, $f$-plane geostrophic estimate, baseline U-Net (E0), and final U-Net (C3). Rows: eastward velocity $u$, northward velocity $v$ (shared color scale per row), and the magnitude of the vector error; the number in each error panel is the RMSE of speed for this patch. The U-Nets shown are the seed-0 runs of each configuration.}
\label{fig:predictions}
\end{figure}

Figure~\ref{fig:predictions} shows a representative test patch. The geostrophic estimate reproduces the large-scale pattern of $u$ and $v$ but contains small-scale structure that is absent from the 53\,m velocity, and the baseline U-Net produces an overly smooth field. The final model reproduces both the large-scale flow and the filaments; its vector error is smaller and spatially more uniform (RMSE of speed 0.038\,m/s for this patch, compared with 0.058\,m/s for geostrophy and 0.060\,m/s for the baseline).

\subsection{Vorticity and Divergence Reconstruction}
\label{sec:kinematic_results}

\begin{table}[h]
\centering
\caption{Vorticity and divergence errors on the test block. RMSEs are of $\zeta/f$ and $\delta/f$ (dimensionless, $f_0=-9.95\times10^{-5}$\,s$^{-1}$); for reference, the standard deviations of the true fields in the test block are 0.266 ($\zeta/f$) and 0.090 ($\delta/f$). U-Net values are means over three seeds.}
\label{tab:kinematic_errors}
\setlength{\tabcolsep}{5pt}
\begin{tabular}{lcccc}
\toprule
 & \multicolumn{2}{c}{Relative vorticity} & \multicolumn{2}{c}{Horizontal divergence} \\
Model & RMSE($\zeta/f$) & $r_\zeta$ & RMSE($\delta/f$) & $r_\delta$ \\
\midrule
Geostrophic estimate & 0.208 & 0.749 & 0.090$^\dagger$ & 0 \\
Baseline U-Net (E0) & 0.177 & 0.756 & 0.094 & 0.109 \\
E4 (augmentation, departure from geostrophy) & 0.119 & 0.896 & 0.082 & 0.394 \\
Final model (C3) & \textbf{0.097} & \textbf{0.931} & \textbf{0.067} & \textbf{0.661} \\
\bottomrule
\multicolumn{5}{l}{\footnotesize $^\dagger$Equals the standard deviation of the true $\delta/f$, since $\nabla\cdot\mathbf{u}_g=0$.}
\end{tabular}
\end{table}

The geostrophic estimate captures most of the velocity itself (its components correlate with the truth at $r=0.93$ over the full record), but its vorticity is noisy ($r_\zeta=0.749$, Table~\ref{tab:kinematic_errors}), because the second derivatives in $\zeta_g=(g/f)\nabla^2\eta$ amplify small-scale SSH signals. Part of this is tidal: on a random sample of 300 patches from the full record, computing $\zeta_g$ from $\eta$ averaged over 25 hours, which removes most of the semidiurnal and diurnal tides, lowers its RMSE from 0.27 to 0.19 and raises its correlation from 0.59 to 0.65. The final model reproduces vorticity far better ($r_\zeta=0.931$) and recovers a large part of the divergence ($r_\delta=0.661$), which the geostrophic estimate cannot represent at all. Only about 37\,\% of the divergence variance at 53\,m is at periods longer than 13\,h (Section~\ref{sec:data}), whereas the final model reproduces about 44\,\% of it ($r_\delta^2=0.44$), so the network also infers part of the higher-frequency, internal-wave divergence from the inputs, presumably from the SSH signature of internal tides. Relative to the variability of each field, vorticity is reconstructed better than divergence: the RMSE is 37\,\% of the standard deviation for $\zeta/f$ and 75\,\% for $\delta/f$.

\begin{figure}[t]
\centering
\includegraphics[width=\textwidth]{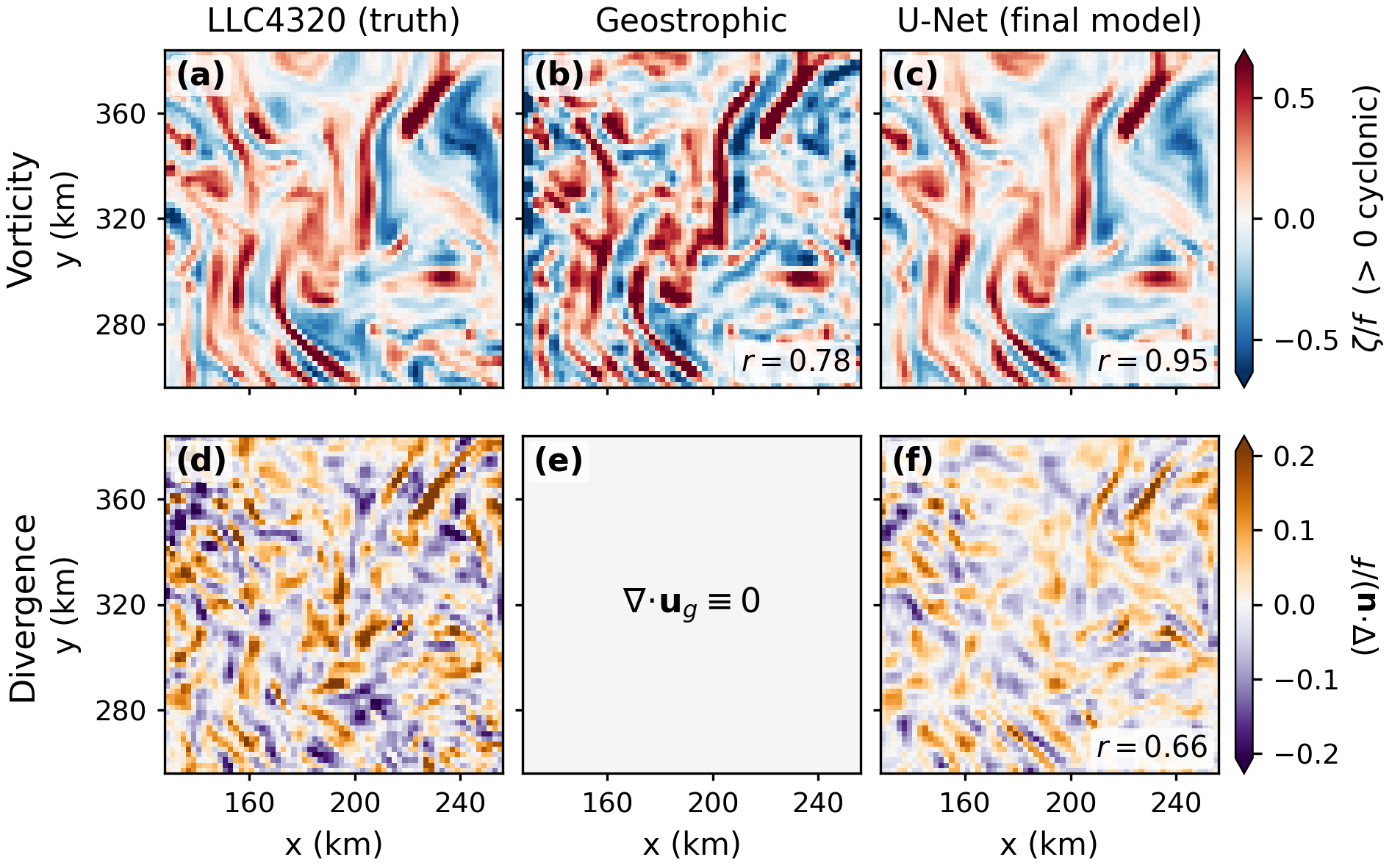}
\caption{Relative vorticity $\zeta/f$ (top) and horizontal divergence $\delta/f$ (bottom) on a representative test patch ($128\times128$\,km; 50\,km scale bar in (d)). Left: LLC4320 truth; center: $f$-plane geostrophic estimate from SSH, whose divergence is zero by construction; right: final model. $f_0<0$, so $\zeta/f>0$ (red) is cyclonic. The correlation with the truth for this patch is given in each panel. The patch is the one whose $r_\zeta$ for the final model (seed-0 run) equals the test-block median.}
\label{fig:kinematic_diagnostics}
\end{figure}

Figure~\ref{fig:kinematic_diagnostics} shows the test patch whose correlation $r_\zeta$ for the final model equals the median over the test block (0.946; interquartile range 0.905--0.963). The geostrophic vorticity of this patch contains much small-scale noise ($r_\zeta=0.78$), while the final model reproduces the eddies and filaments ($r_\zeta=0.95$) and much of the banded divergence pattern ($r_\delta=0.66$), whose wave-like structure is consistent with internal gravity waves.

\subsection{Velocity at the Surface}
\label{sec:surface}

To test the approach for the surface velocity, the quantity most often estimated from altimetry, we trained the baseline and final configurations with the velocity of the top model level (0.5\,m) as target, without changing any other setting, together with the final configuration without wind stress and without SST (Table~\ref{tab:surface}).

\begin{table}[h]
\centering
\caption{Surface velocity (0.5\,m): test and validation RMSE of speed (m/s; mean $\pm$ standard deviation over three seeds), test RMSE of each component, direction error, and correlations with the true vorticity and divergence. All U-Nets use the configurations of Section~\ref{sec:generalization} without retuning.}
\label{tab:surface}
\setlength{\tabcolsep}{3.5pt}
\begin{tabular}{lccccccc}
\toprule
Model & Test RMSE speed & Val.\ RMSE speed & RMSE$_u$ & RMSE$_v$ & Dir.\ error & $r_\zeta$ & $r_\delta$ \\
\midrule
Geostrophic estimate & 0.1249 & 0.1220 & 0.1389 & 0.1114 & 35.1$^\circ$ & 0.762 & 0 \\
Baseline U-Net (as E0) & 0.0718 $\pm$ 0.0013 & 0.0701 & 0.0674 & 0.0690 & 16.8$^\circ$ & 0.807 & 0.307 \\
Final U-Net (as C3) & 0.0511 $\pm$ 0.0009 & 0.0488 & 0.0488 & 0.0494 & 12.1$^\circ$ & 0.973 & 0.874 \\
\quad without wind stress & 0.0803 $\pm$ 0.0014 & 0.0763 & 0.0792 & 0.0836 & 20.7$^\circ$ & 0.907 & 0.605 \\
\quad without SST & 0.0492 $\pm$ 0.0006 & 0.0487 & 0.0482 & 0.0489 & 12.2$^\circ$ & 0.967 & 0.841 \\
\bottomrule
\end{tabular}
\end{table}

Geostrophy is much less accurate at the surface than at 53\,m: its RMSE of speed is 0.125\,m/s (0.069\,m/s at 53\,m), its direction error 35$^\circ$, and it explains only 59\,\% and 77\,\% of the variance of $u$ and $v$. The surface velocity contains wind-driven Ekman and near-inertial currents, which are not geostrophic. The baseline U-Net is better than geostrophy at the surface (0.0718\,m/s) but again reaches its lowest validation error after two to four epochs. The final configuration reduces the RMSE of speed to $0.0511\pm0.0009$\,m/s (RMSE$_u=0.0488$, RMSE$_v=0.0494$\,m/s; $R^2_u=0.949$, $R^2_v=0.955$), 59\,\% below geostrophy and close to its error at 53\,m. It reproduces the surface vorticity and divergence with $r_\zeta=0.973$ and $r_\delta=0.874$, compared with 0.762 and 0 for geostrophy; the RMSE is 23\,\% of the standard deviation for $\zeta/f$ and 49\,\% for $\delta/f$.

Wind stress is essential at the surface: without it, the RMSE of speed increases by 57\,\% (to 0.0803\,m/s), compared with 20\,\% at 53\,m (0.0486 to 0.0581\,m/s; Table~\ref{tab:ablation}). Without SST, the test error is slightly lower (0.0492\,m/s) and the validation error is the same (0.0487 versus 0.0488\,m/s), so, as at 53\,m, SST does not improve the test skill.

\section{Joint Statistics of Vorticity and Strain}
\label{sec:validation_results}

\begin{figure}[t]
\centering
\includegraphics[width=\textwidth]{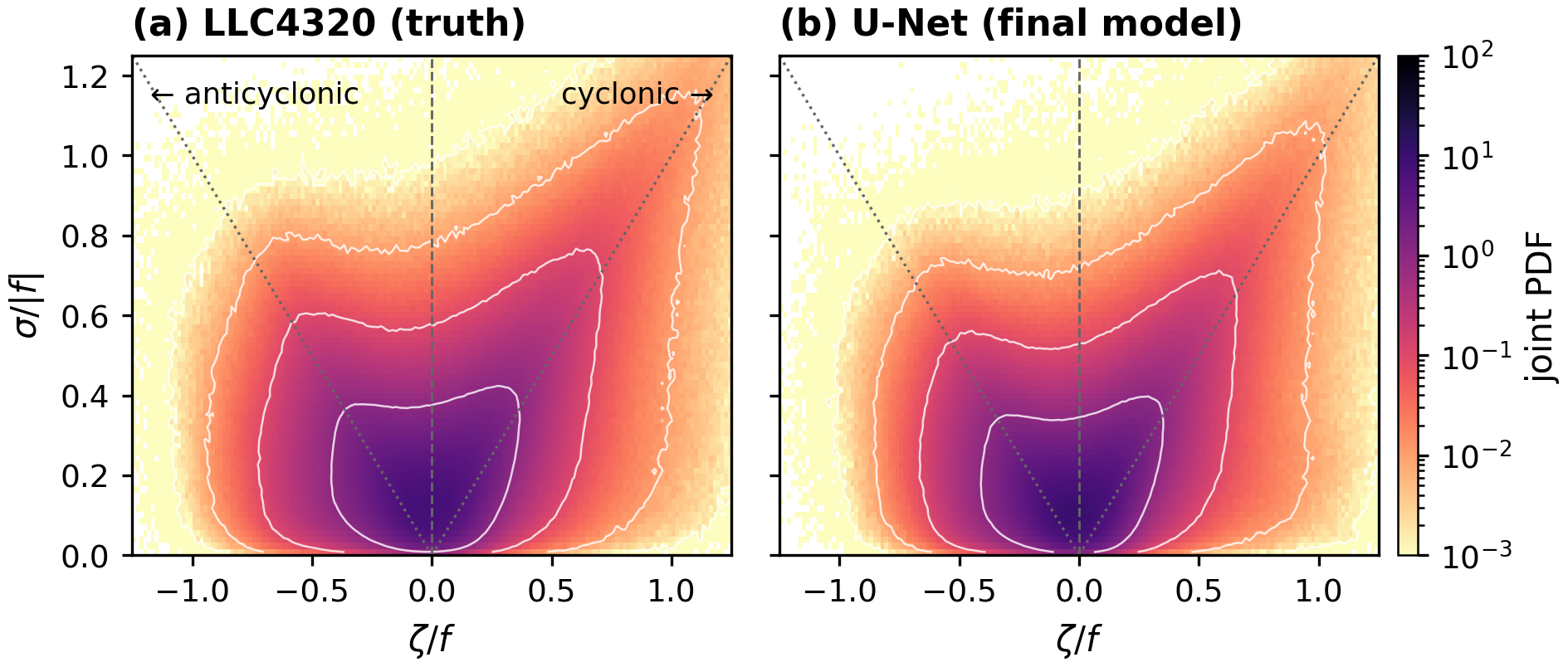}
\caption{Joint PDF of $\zeta/f$ and $\sigma/|f|$ over all test pixels: (a)~LLC4320 truth; (b)~final model (seed-0 run). $f_0<0$, so $\zeta/f>0$ is cyclonic. White contours mark densities of $10^{-3}$, $10^{-2}$, $10^{-1}$, and 1; dotted lines mark $\sigma/|f|=|\zeta/f|$. The PDF is normalized by the total number of pixels; the plotted range contains 99.9\,\% of them.}
\label{fig:vorticity_strain_pdf}
\end{figure}

Figure~\ref{fig:vorticity_strain_pdf} shows the joint probability density function (PDF) of $\zeta/f$ and the normalized strain rate $\sigma/|f|$ in the test block for the LLC4320 truth and the final model, following \citet{Balwada2021}, who use the same normalization by a constant $f_0$ ($\zeta/f_0$, $\sigma/|f_0|$). Here
\begin{equation}
    \sigma = \sqrt{\left(\frac{\partial u}{\partial x} - \frac{\partial v}{\partial y}\right)^{\!2}
                  + \left(\frac{\partial u}{\partial y} + \frac{\partial v}{\partial x}\right)^{\!2}} \;\geq 0 .
\end{equation}

The true distribution (Fig.~\ref{fig:vorticity_strain_pdf}a) is skewed toward cyclonic vorticity (skewness of $\zeta/f$ = 0.35) and extends along the lines $\sigma\approx|\zeta|$, with a longer, high-strain tail on the cyclonic side that reaches $\zeta/f\approx1$. A cyclonic skewness and strain-dominated fronts ($\sigma\gtrsim|\zeta|$) are characteristic of submesoscale fronts and filaments \citep{McWilliams2016,Balwada2021}. The predicted PDF (panel~b) reproduces the shape of the distribution, the cyclonic high-strain tail, and the cyclonic skewness, which it slightly overestimates (0.42). It is narrower than the truth: the standard deviation of $\zeta/f$ is 0.240 instead of 0.266, and strain rates above $0.5|f|$ occur at 3.7\,\% of the pixels instead of 5.8\,\%, consistent with smoothing of the sharpest fronts.

\section{Spectral Energy Flux}
\label{sec:flux}

Beyond pointwise errors, a velocity estimate should reproduce how kinetic energy is transferred between scales. \citet{Scott2005} computed this transfer from altimetric geostrophic velocities and found an inverse cascade toward large scales. We compare the spectral kinetic-energy flux of the true velocity with that of the geostrophic and predicted velocities.

\subsection{Definition}

For a horizontal velocity field $\mathbf{u}=(u,v)$ on the uniform 2\,km grid, let $\hat{\mathbf{u}}(\mathbf{k})$ be its two-dimensional Fourier transform. The kinetic energy at wavenumber $\mathbf{k}$ is $\frac{1}{2}|\hat{\mathbf{u}}(\mathbf{k})|^2$, and its rate of change due to horizontal advection is the spectral transfer
\begin{equation}
	T(\mathbf{k}) = -\mathrm{Re}\left[\hat{\mathbf{u}}^*(\mathbf{k})\cdot\widehat{(\mathbf{u}\cdot\nabla)\mathbf{u}}(\mathbf{k})\right].
	\label{eq:transfer}
\end{equation}
Advection only redistributes energy among wavenumbers, so for a periodic, non-divergent flow the transfer sums to zero. The spectral flux through the isotropic wavenumber $K$,
\begin{equation}
	\Pi(K) = \sum_{|\mathbf{k}|\geq K} T(\mathbf{k}),
	\label{eq:flux}
\end{equation}
is the rate at which energy is transferred from wavenumbers below $K$ to wavenumbers above $K$, that is, from scales larger than the wavelength $2\pi/K$ to smaller scales. $\Pi>0$ indicates a forward cascade toward small scales and $\Pi<0$ an inverse cascade toward large scales. Equation~\eqref{eq:transfer} describes the transfer by horizontal advection at a single depth; vertical advection is not included.

\subsection{Computation}

For each hour of the test period, we stitch the six tiles into the $192\times128$-cell (384\,km $\times$ 256\,km) domain. The U-Net is fully convolutional, so we apply it to the whole domain at once; this avoids artificial discontinuities at tile boundaries, and the final model reaches a RMSE of speed of 0.0473\,m/s at 53\,m and 0.0502\,m/s at the surface on the whole domain (mean over seeds). The advection term is computed with centered differences, and both the velocity and the advection term are multiplied by a two-dimensional Tukey window (taper fraction 0.25) before the Fourier transform. Modes are binned in isotropic wavenumber with a bin width of $2\pi/(256\,\mathrm{km})$, and $T$, $\Pi$, and the kinetic-energy spectrum $E(K)$ are averaged over the 1{,}992 test hours. Because the domain is open and the flow divergent, the transfer does not sum exactly to zero, and the residual affects the domain-scale wavenumbers. Equation~\eqref{eq:flux} sums from the small-scale end, so $\Pi$ at smaller scales does not depend on the domain-scale modes; we show wavelengths up to 85\,km. Changing the taper fraction to 0.1 or 0.5 changes the magnitude of $\Pi$, but the ratios of the geostrophic and U-Net extrema to the true extrema change by at most 0.05.

\paragraph{Robustness.} For a divergent flow, the advective form of Eq.~\eqref{eq:transfer} does not conserve energy exactly. We therefore repeated the analysis on 120 evenly spaced test hours with the skew-symmetric form $(\mathbf{u}\cdot\nabla)\mathbf{u}+\frac{1}{2}\mathbf{u}(\nabla\cdot\mathbf{u})$, which conserves energy for periodic flows of any divergence. The two forms coincide for the non-divergent geostrophic flow. The magnitudes of $\Pi$ and the wavelength at which the true flux changes sign depend on the form (20--34\,km at 53\,m and 22--30\,km at the surface). The conclusions below hold for both forms.

\begin{figure}[t]
\centering
\includegraphics[width=\textwidth]{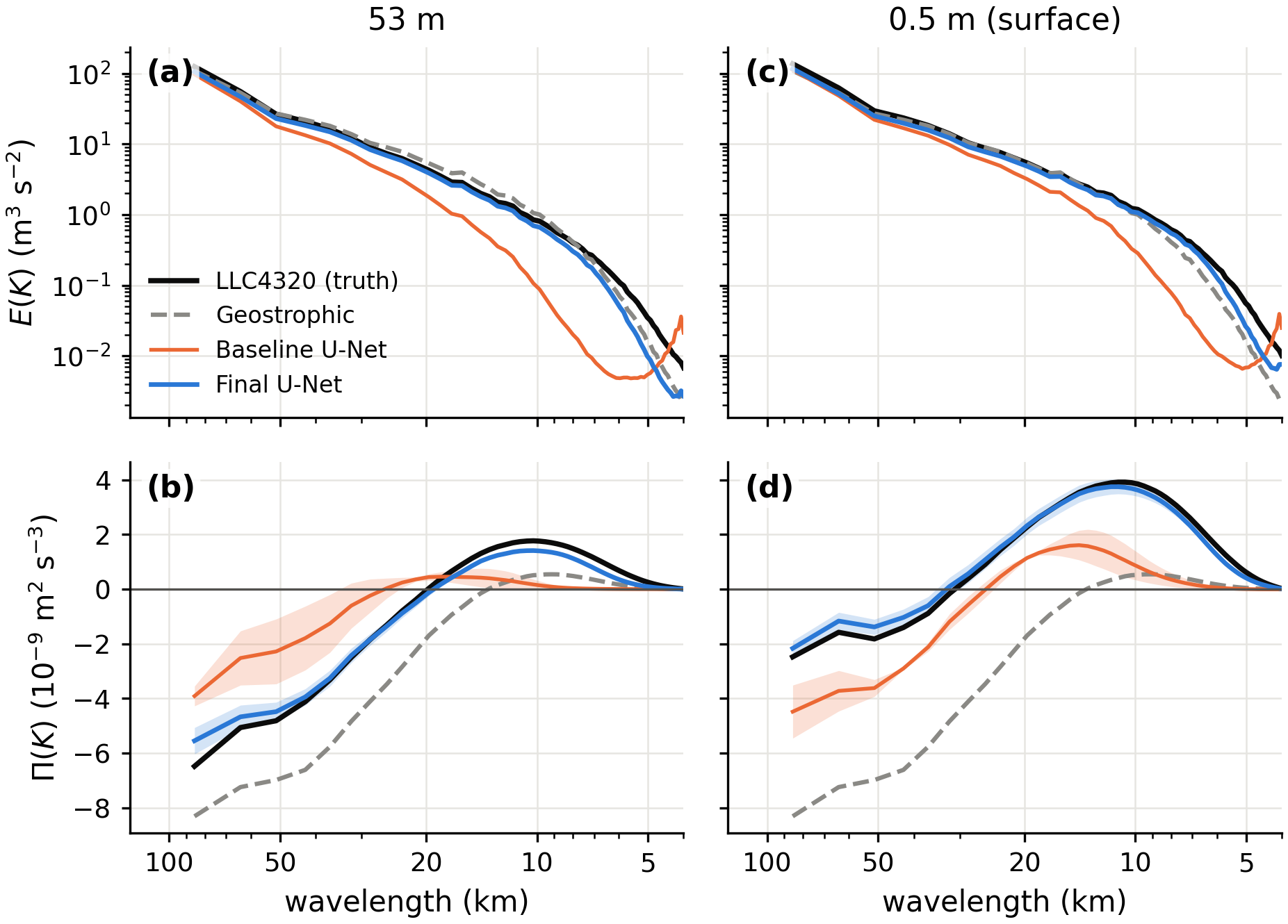}
\caption{Kinetic-energy spectrum $E(K)$ (top) and spectral kinetic-energy flux $\Pi(K)$ (bottom) as functions of wavelength $2\pi/K$, averaged over the test period: (a, b)~at 53\,m; (c, d)~at the surface. $\Pi>0$ is a forward cascade toward small scales, $\Pi<0$ an inverse cascade toward large scales. The geostrophic estimate is computed from the same SSH at both depths. Shading: one standard deviation over three seeds.}
\label{fig:spectral_flux}
\end{figure}

\subsection{Results}

Figure~\ref{fig:spectral_flux} and Table~\ref{tab:flux} summarize the spectra and fluxes. At both depths the true velocity shows a dual cascade: an inverse cascade at large scales and a forward cascade at small scales, which in the advective form changes sign at about 20\,km (53\,m) and 30\,km (surface) and peaks near 10--11\,km. In both forms the forward cascade is stronger at the surface than at 53\,m.

\begin{table}[h]
\centering
\caption{Spectral flux on the test period, advective form of Eq.~\eqref{eq:transfer}: largest inverse flux (minimum of $\Pi$, at the 85\,km wavelength), wavelength at which $\Pi$ changes sign, and largest forward flux (maximum of $\Pi$) with its wavelength. $\Pi$ in $10^{-9}$\,m$^2$\,s$^{-3}$; U-Net values are means $\pm$ standard deviations over three seeds.}
\label{tab:flux}
\setlength{\tabcolsep}{4pt}
\begin{tabular}{lcccccc}
\toprule
 & \multicolumn{3}{c}{53\,m} & \multicolumn{3}{c}{Surface (0.5\,m)} \\
 & min $\Pi$ & sign change & max $\Pi$ & min $\Pi$ & sign change & max $\Pi$ \\
\midrule
LLC4320 (truth) & $-6.5$ & 20\,km & 1.77 (10\,km) & $-2.5$ & 30\,km & 3.92 (11\,km) \\
Geostrophic estimate & $-8.3$ & 14\,km & 0.54 (9\,km) & $-8.3$ & 14\,km & 0.54 (9\,km) \\
Baseline U-Net & $-3.9\pm0.4$ & 27\,km & $0.46\pm0.17$ (18\,km) & $-4.5\pm1.0$ & 24\,km & $1.61\pm0.54$ (14\,km) \\
Final U-Net & $-5.6\pm0.5$ & 19\,km & $1.41\pm0.05$ (10\,km) & $-2.2\pm0.3$ & 34\,km & $3.75\pm0.28$ (12\,km) \\
\bottomrule
\end{tabular}
\end{table}

The geostrophic estimate misrepresents this cascade. Its forward cascade is only 31\,\% of the true one at 53\,m and 14\,\% at the surface (18\,\% and 11\,\% in the skew-symmetric form). Its inverse cascade is at least as strong as the true one (1.0 to 3.4 times as strong, depending on depth and form) and changes sign at a smaller scale (14\,km). This happens although its kinetic energy at 10--20\,km wavelength is within 15\,\% of, or at 53\,m even 25\,\% higher than, the true one (Fig.~\ref{fig:spectral_flux}a, c). The $f$-plane geostrophic flow is non-divergent and lacks the ageostrophic circulation of fronts and filaments; its flux is consistent with two-dimensional, non-divergent dynamics, in which energy is transferred mainly toward large scales. The baseline U-Net, whose small-scale energy is too low, also underestimates the forward cascade. The final U-Net reproduces the dual cascade at both depths: its forward cascade reaches 80\,\% (53\,m) and 96\,\% (surface) of the true maximum in the advective form and 77\,\% and 95\,\% in the skew-symmetric form, and its sign change lies within 5\,km of the true one in both forms.

\section{Discussion}
\label{sec:discussion}

\paragraph{What makes the network generalize.} The training region contains only six fixed subdomains and ten months of hourly fields, that is, relatively few independent flow configurations. The two augmentations increase the diversity of the training samples without changing the physics: rotations preserve the $f$-plane dynamics, including geostrophic balance, while changing the orientation of the flow and of the wind stress, and random crops present the same fields in many more windows. Predicting the departure from geostrophy leaves the network to learn smaller residuals, and the vorticity and divergence penalties make it attend to the spatial derivatives of the velocity and reduces spatial discontinuous artifacts, which a pointwise loss weights only weakly. The same configuration applies to the surface velocity, where geostrophy is much less accurate because of wind-driven currents. The model performance is further shown in the accurate recreation of the spectral transfer of kinetic energy, where geostrophic approximation fails as expected at small scales by underestimating forward cascade and overestimating inverse cascade.

This work examined the inference of ocean velocity at 53\,m depth and at the surface from SSH snapshots and surface covariates with a convolutional encoder--decoder, using hourly LLC4320 fields from the pre-SWOT Cape Basin region. The main findings are as follows.

\paragraph{Generalization.} Evaluated on a held-out period, a standard U-Net is no better than geostrophy. Rotation and crop augmentation, predicting the departure from geostrophy, vorticity and divergence penalties, and a base width of 128 reduce the test RMSE of speed to $0.0486\pm0.0007$\,m/s (RMSE$_u$ = 0.046\,m/s, RMSE$_v$ = 0.046\,m/s, $R^2\approx0.95$), 29\,\% below geostrophy.

\paragraph{Inputs and training.} SSH gradients are at least as good inputs as raw SSH for the RMSE of speed and better for vorticity; wind stress is the most useful covariate; the effect of SST differs between the validation and test periods; capacity saturates at a base width of 128; and MSE and MAE perform equally, while the vorticity and divergence penalties clearly help.

\paragraph{Surface velocity.} Trained for the surface velocity without retuning, the final configuration reaches an RMSE of speed of $0.0511\pm0.0009$\,m/s, 59\,\% below geostrophy (0.125\,m/s); wind stress is essential at the surface (57\,\% higher error without it).

\paragraph{Spectral energy flux.} The true flow has an inverse cascade at large scales and a forward cascade at scales below a few tens of kilometers. Geostrophy captures at most a third of the forward cascade, whereas the final U-Net reproduces 77--96\,\% of it at the correct scales; these conclusions hold for both forms of the nonlinear term.

\paragraph{Kinematic quality.} The final model reproduces vorticity with $r_\zeta=0.93$ (0.75 for geostrophy), recovers much of the divergence ($r_\delta=0.66$), which at 53\,m is dominated by internal gravity waves and which geostrophy cannot represent, and reproduces the joint distribution of vorticity and strain, including its cyclonic skewness, with somewhat too few high-strain values.

\section{Conclusion}
\label{sec:conclusion}
Inferring velocity from sea surface height alone is mathematically underdetermined in the nonlinear regimes revealed by SWOT. Although the conventional formulation does not yield a unique solution, deep neural networks may exploit information embedded in a single SSH snapshot to constrain the velocity field. The mechanisms remain unclear, but spatial patterns and spectral phase relationships among different physical processes may provide information that conventional methods do not fully exploit and that is difficult to discern visually. Our study here, together with previous published work, demonstrates the potential of this approach. Establishing its reliability, however, requires validation against truly independent observations, which is the focus of our ongoing work. 

\subsection*{Acknowledgments}

This work is supported by NASA (80NSSC26K0212, 80NSSC25K7643) and the Summer Undergraduate Research Fellowships (SURF) at Jet Propulsion Laboratory, California Institution of Technology. Portions of this research were conducted with advanced computing resources provided by Texas A\&M High Performance Research Computing (HPRC).

\subsection*{Data and Code Availability}

The LLC4320 pre-SWOT Cape Basin data used in this study are publicly available from NASA PO.DAAC (\nolinkurl{MITgcm_LLC4320_Pre-SWOT_JPL_L4_CapeBasin_v1.0}; \url{https://doi.org/10.5067/PRESW-CBJ10}). Training code, experiment configurations, and evaluation scripts are available at \url{https://github.com/odsl/ssh2uv}.

\appendix

\section{All Temporal-Split Experiments}
\label{app:all_runs}

Table~\ref{tab:all_runs} lists all experiments trained with the temporal split. Unless stated otherwise, they use SSH gradients, SST, and wind stress as inputs, a base width of 64, MSE loss, and a maximum of 60 epochs with a patience of 10.

\begin{table}[h]
\centering
\caption{All temporal-split experiments: test and validation RMSE of speed (m/s; mean $\pm$ standard deviation over three seeds) and test correlations. ``Residual'': departure from geostrophy as target; ``pen.'': vorticity/divergence penalties ($w_\zeta=w_\delta=0.5$ unless stated); ``long'': up to 100 epochs, patience 20.}
\label{tab:all_runs}
\setlength{\tabcolsep}{3pt}
\footnotesize
\begin{tabular}{llcccc}
\toprule
 & Configuration & Test RMSE speed & Val.\ RMSE speed & $r_\zeta$ & $r_\delta$ \\
\midrule
E0 & $B{=}64$, no augmentation & 0.0681 $\pm$ 0.0021 & 0.0740 & 0.756 & 0.109 \\
E1 & E0 + rotations & 0.0604 $\pm$ 0.0018 & 0.0632 & 0.812 & 0.169 \\
E2 & E0 + random crops & 0.0627 $\pm$ 0.0013 & 0.0660 & 0.791 & 0.167 \\
E3 & E0 + rotations + crops & 0.0562 $\pm$ 0.0007 & 0.0577 & 0.861 & 0.302 \\
E4 & E3, residual & 0.0556 $\pm$ 0.0006 & 0.0545 & 0.896 & 0.394 \\
E5 & E3, $\mathbf{u}_g$ as additional input & 0.0573 $\pm$ 0.0004 & 0.0583 & 0.853 & 0.290 \\
E6 & E3, raw SSH instead of gradients & 0.0617 $\pm$ 0.0015 & 0.0669 & 0.847 & 0.218 \\
\midrule
R1a & E4, $B{=}128$ & 0.0530 $\pm$ 0.0022 & 0.0520 & 0.913 & 0.518 \\
R1b & E4, $B{=}192$ & 0.0529 $\pm$ 0.0004 & 0.0516 & 0.915 & 0.530 \\
R2a & E4 without SST & 0.0520 $\pm$ 0.0007 & 0.0568 & 0.892 & 0.434 \\
R2b & E4 without wind stress & 0.0589 $\pm$ 0.0005 & 0.0583 & 0.861 & 0.322 \\
R2c & E4 with SSH gradients only & 0.0574 $\pm$ 0.0015 & 0.0623 & 0.832 & 0.202 \\
R3a & E4, MAE loss & 0.0556 $\pm$ 0.0011 & 0.0547 & 0.894 & 0.391 \\
R3b & E4 + pen. & 0.0516 $\pm$ 0.0015 & 0.0513 & 0.921 & 0.627 \\
R4 & E4, long & 0.0545 $\pm$ 0.0010 & 0.0543 & 0.905 & 0.457 \\
R5a & E4, $\mathbf{u}_g$ from SSH smoothed at 4\,km & 0.0560 $\pm$ 0.0011 & 0.0554 & 0.892 & 0.384 \\
R5b & E4, $\mathbf{u}_g$ from SSH smoothed at 8\,km & 0.0556 $\pm$ 0.0020 & 0.0553 & 0.887 & 0.370 \\
\midrule
C1 & R3b, $B{=}128$ & 0.0496 $\pm$ 0.0007 & 0.0496 & 0.930 & 0.661 \\
C2 & R3b, $B{=}192$ & 0.0500 $\pm$ 0.0008 & 0.0490 & 0.933 & 0.675 \\
C3 & C1, long (final model) & 0.0486 $\pm$ 0.0007 & 0.0490 & 0.931 & 0.661 \\
C4 & C1, $w_\zeta=w_\delta=1$ & 0.0502 $\pm$ 0.0008 & 0.0493 & 0.933 & 0.675 \\
C5 & C1 without SST & 0.0481 $\pm$ 0.0004 & 0.0528 & 0.917 & 0.604 \\
\midrule
S-E0 & E0, surface velocity & 0.0718 $\pm$ 0.0013 & 0.0701 & 0.807 & 0.307 \\
S-C3 & C3, surface velocity & 0.0511 $\pm$ 0.0009 & 0.0488 & 0.973 & 0.874 \\
S-C3a & S-C3 without wind stress & 0.0803 $\pm$ 0.0014 & 0.0763 & 0.907 & 0.605 \\
S-C3b & S-C3 without SST & 0.0492 $\pm$ 0.0006 & 0.0487 & 0.967 & 0.841 \\
\bottomrule
\end{tabular}
\end{table}

The ablations of Section~\ref{sec:ablation} (raw SSH and SSH gradients with each covariate set, base widths 64 and 192, and the four losses) are variants of C3 and are reported in Tables~\ref{tab:ablation}--\ref{tab:loss_ablation}.

\section{Random Versus Temporal Split}
\label{sec:leakage}

\begin{figure}[!htb]
\centering
\includegraphics[width=\textwidth]{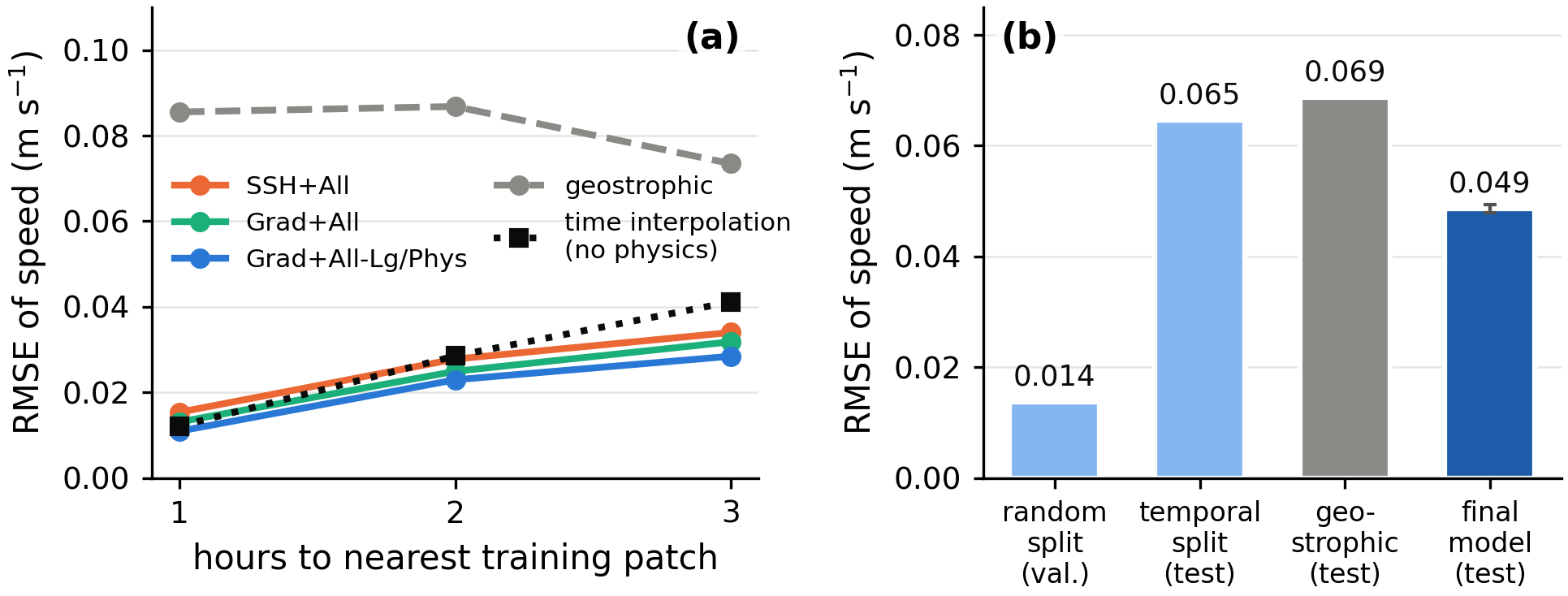}
\caption{Random versus temporal split. (a)~RMSE of speed on the random-split validation set as a function of the time to the nearest training patch of the same subdomain, for three U-Nets trained on the random split (SSH+All: raw SSH, SST, and wind stress, $B{=}192$; Grad+All: SSH gradients, SST, and wind stress, $B{=}192$; Grad+All-Lg/Phys: the same at $B{=}384$ with vorticity and divergence penalties), the $f$-plane geostrophic estimate, and an interpolation in time between the neighboring training patches. (b)~The Grad+All configuration on the random-split validation set and, trained on the temporal split, on the test period, with geostrophy and the final model (mean and standard deviation over three seeds) on the same test period.}
\label{fig:leakage}
\end{figure}

For comparison, we trained U-Nets on a random 80/20 split of all 61{,}176 patches (seed 42; 48{,}941 training and 12{,}235 validation patches), with the validation set also used for checkpoint selection. These models use raw SSH or SSH gradients with SST and wind stress, $B{=}192$ or 384, an MAE loss (in one variant MSE, in another MAE with vorticity and divergence penalties) with a cosine direction term (weight 0.04), and normalization statistics from 10{,}000 patches of the full record. In this split, 96\,\% of the validation patches have a training patch from the same subdomain one hour earlier or later, and the velocity changes little in one hour (RMSE of speed 0.024\,m/s, rising to 0.044\,m/s after two hours and 0.087\,m/s after six). Predicting each validation patch by linear interpolation in time between the nearest training patches of the same subdomain, without any model, gives an RMSE of speed of 0.0132\,m/s, lower than that of the SSH-input (0.0160\,m/s) and gradient-input (0.0138\,m/s) U-Nets at $B{=}192$. The error of the trained models increases by a factor of 1.8--2.0 when the nearest training patch is two hours away instead of one, whereas that of geostrophy does not (Fig.~\ref{fig:leakage}a). Trained on the temporal split, the gradient-input configuration at $B{=}192$ reaches 0.0646\,m/s on the test period, 4.7 times its random-split score and close to geostrophy (0.0687\,m/s; Fig.~\ref{fig:leakage}b); its validation loss is lowest after six epochs.

\clearpage 
\bibliographystyle{unsrtnat}
\bibliography{references}

\end{document}